\documentclass[aps,prd,onecolumn,nofootinbib,superscriptaddress]{revtex4}
\pdfoutput=1
\usepackage{graphicx}
\usepackage{amsmath}
\usepackage{amsfonts}
\usepackage{amssymb,ulem}
\usepackage{color}%
\usepackage{dcolumn}

\usepackage{MnSymbol,wasysym}
\usepackage{braket}
\usepackage{eurosym}
\usepackage{calrsfs}
\usepackage{multirow}
\usepackage{float}
\usepackage[usenames,dvipsnames,svgnames]{xcolor}

\newcommand{\RNum}[1]{\uppercase\expandafter{\romannumeral #1\relax}}
\usepackage[colorlinks=true,linkcolor=blue,urlcolor=blue,filecolor=black,citecolor=red,pdfstartview=FitV,pdftitle={},pdfsubject={},pdfkeywords={},pdfpagemode=None,bookmarksopen=true]{hyperref}
\usepackage[title]{appendix}
\makeatletter
\newcommand*{\rom}[1]{\expandafter\@slowromancap\romannumeral #1@}
\makeatother

\begin{document}
\baselineskip=0.5 cm

\title{Gravitational odd-parity perturbation of a Horndeski hairy black hole: quasinormal mode and parameter constraint}

\author{Zhen-Hao Yang}
\affiliation{Center for Gravitation and Cosmology, College of Physical Science and Technology, Yangzhou University, Yangzhou, 225009, China}
\author{Yun-He Lei}
\affiliation{Center for Gravitation and Cosmology, College of Physical Science and Technology, Yangzhou University, Yangzhou, 225009, China}
\author{Xiao-Mei Kuang}
\email{xmeikuang@yzu.edu.cn (corresponding author)}
\affiliation{Center for Gravitation and Cosmology, College of Physical Science and Technology, Yangzhou University, Yangzhou, 225009, China}
\author{Bin Wang}
\affiliation{Center for Gravitation and Cosmology, College of Physical Science and Technology, Yangzhou University, Yangzhou, 225009, China}
\affiliation{Shanghai Frontier Science Center for gravitational waves Detection, Shanghai Jiao Tong University, Shanghai 200240, China}

\begin{abstract}
\baselineskip=0.5 cm
During the binary black hole coalescence, gravitational waves emitted at the ringdown stage can be well described by black hole perturbation theory, where the quasinormal modes (QNMs)  become the important ingredient in modeling the pattern waveform. In general relativity (GR), the QNMs can be obtained from solving the Regge-Wheeler (RW) equation of a non-rotating black hole. While in Horndeski gravity, the isospectrality between the odd and even parity perturbations is broken due to the scalar field, but the odd  perturbation equation can be simplified into a modified RW equation from the perturbed action. In this paper, we propose a new auxiliary field and tortoise coordinate to refine the modified RW equation in Horndeski gravity, and calculate the QNMs frequencies of the odd perturbation of a specific hairy black hole. It is found that this proposal not only cures the superluminal propagation addressed in the previous literature, but also hold the original QNM spectrum of the odd perturbation.  Moreover, our results indicate that such a Horndeski hairy black hole is stable under the odd perturbation, which is also verified by the time evolution of the perturbation. In particular, in contrary to GR, the  modes with $\ell=2$ can decay faster than  modes with $\ell>2$ for a certain range of the Horndeski hair, and the link between the null geodesics and QNM for the odd perturbation in the current theory is violated. Then, we use the ringdown QNMs to preliminarily  investigate the signal-to-noise-ratio (SNR) rescaled measurement error of the Horndeski hair. We obtain significant effects of the angular momentum and overtone on the error bound of the hair parameter. We hope that our findings could inspire more theoretical and phenomenological work on the test of the no-hair theorem of black hole using gravitational wave physics.
\end{abstract}

\maketitle
\newpage
\tableofcontents

\section{Introduction}
Testifying the no-hair theorem of black hole \cite{PhysRev.164.1776, PhysRevD.5.1239, BEKENSTEIN1974535, Hawking1972} and its extensions to other theories of gravity  is important to test the nature of gravity and more fundamental physics, which inevitably requires us to explore the strong gravity regime. 
Potential ways of seeking the black hole hair are to investigate the imprints of hair on the gravitaional wave (GW) and on the shadow from the black hole. 
Recent GW observations from the merger of black holes or other compact objects \cite{LIGOScientific:2016aoc,LIGOScientific:2018mvr,LIGOScientific:2020aai} and black hole shadows  from supermassive black holes \cite{EventHorizonTelescope:2019dse,EventHorizonTelescope:2022xnr} open a new era of exploring physics about the strong gravity regime, which triggers extensive investigation about the prints of hair on the GW and shadows. The studies on testing the no-hair theorem with Event Horizon Telescope (EHT) observations can refer to, for examples, the literature \cite{Broderick:2013rlq,Psaltis:2015uza,Khodadi:2020jij,Khodadi:2021gbc,Tang:2022uwi}, but in this scenario, a universal challenge is achieving high precision in detecting the print of hair on black hole shadow. This is because the hair should directly affect the orbits very near the black hole horizon, but the photon ring still lies far away from the horizon. GWs generated during the extreme mass ratio inspiral (EMRI) and ringdown stage in binary black hole coalescence are potential tools to seek the black hole hair \cite{Thrane:2017lqn,Isi:2019aib}. During the stage of EMRI the hair will influence the inspiraling orbits and energy radiation \cite{Rodriguez:2011aa,Kuntz:2020yow,Maselli:2020zgv,Collodel:2021jwi,Zi:2021pdp,Guo:2022euk,Liang:2022gdk,Zhang:2022rfr,Zi:2023omh,Guo:2023mhq}, while at the ringdown stage, the hair will characterize the ringdown GW wave \cite{Kamaretsos:2011um, Gossan:2011ha,Meidam:2014jpa,Shi:2019hqa,Bhagwat:2019dtm}. Therefore, it is important to concern both cases to completely trace the imprint of black hole hair on the GWs.

Introducing a scalar field into the action of a gravitational sector is a natural way to construct black hole with scalar hair. 
Among the numerous gravity theories, the Horndeski gravity, which contains higher derivatives of a scalar field $\phi$ and metric tensor $g_{\mu\nu}$, has retained lots of attention, since it possesses at most second-order differential field equations so as to be free of Ostrogradski instabilities \cite{Horndeski:1974wa}. Various observational constraints and bounds on Horndeski theories have been explored in \cite{Bellini:2015xja,Bhattacharya:2016naa,Kreisch:2017uet,Hou:2017cjy,SpurioMancini:2019rxy,Allahyari:2020jkn,Ghosh:2023kge} and references therein.
Horndeski gravity is also actively studied in the cosmological and astrophysical communities since it has significant consequences in describing the accelerated expansion and other interesting features \cite{Kobayashi:2019hrl} . Moreover,  since Horndeski theory contains a scalar field, and similar to GR it has  diffeomorphism invariance and second-order field equations, so it also provides a natural  platform to verify the no-hair theorem of black hole. In fact,  hairy black holes in Horndeski gravity have been widely studied, for instance,  the radially
dependent scalar field \cite{Rinaldi:2012vy,Cisterna:2014nua,Feng:2015oea,Sotiriou:2013qea, Miao:2016aol,Kuang:2016edj,Babichev:2016rlq,Benkel:2016rlz,Filios:2018xvy,Cisterna:2018hzf,Giusti:2021sku,Cisterna:2015uya,Anabalon:2013oea,Fontana:2018fof} and the time-dependent scalar field \cite{Babichev:2013cya,Babichev:2017lmw,BenAchour:2018dap,Takahashi:2019oxz,Minamitsuji:2019shy,Arkani-Hamed:2003juy}. The hairy solution with a scalar hair in linear time dependence was initially ruled out due to instability \cite{Khoury:2020aya}. However, another novel solution with a time-dependent scalar hair was recently constructed  \cite{Bakopoulos:2023fmv} and found to be stable \cite{Sirera:2024ghv}. The stability conditions for a similar time-dependent background solution beyond Horndeski theories was also  addressed in \cite{Mironov:2024yqa}. Moreover, it was demonstrated in \cite{Hui:2012qt} ,  that the no-hair theorem cannot hold when a Galileon field is coupled to gravity, but the static spherical solution only admits a trivial Galileon profile. 

An interesting new development in the examination of no-hair theorem in Horndeski theories is that the shift-symmetric Horndeski theory and beyond with the action \cite{Babichev:2017guv}
\begin{eqnarray}\label{eq:action0}
S=&&\int d^4x \sqrt{-g}\big[Q_2+Q_3\Box\phi+Q_4R+Q_{4,\chi}\left((\Box\phi)^2-(\nabla^\mu\nabla^\nu\phi)(\nabla_\mu\nabla_\nu\phi)\right)\nonumber\\
&&+Q_5G_{\mu\nu}\nabla^\mu\nabla^\nu\phi-\frac{1}{6}Q_{5,\chi}\left((\Box\phi)^3-3(\Box\phi)(\nabla^\mu\nabla^\nu\phi)(\nabla_\mu\nabla_\nu\phi)
+2(\nabla_\mu\nabla_\nu\phi)(\nabla^\nu\nabla^\gamma\phi)(\nabla_\gamma\nabla^\mu\phi)\right)\big]
\end{eqnarray}
allow static and asymptotically flat black holes with a nontrivial static scalar field. In the above action, $g$ is the determinant of the metric and $R$ is the Ricci scalar. $\chi=-\partial^\mu\phi\partial_\mu\phi/2$ is the canonical kinetic term, $Q_i~(i=2,3,4,5)$ are arbitrary functions of the scalar field $\phi$ and the kinetic term $\chi$, and $Q_{i,\chi} \equiv \partial Q_{i}/\partial \chi $.
In particular, a static hairy black hole in a specific quartic Horndeski theory, which includes an additional $\log$-term deforming from the standard Schwarzschild black hole,  has been constructed in \cite{Bergliaffa:2021diw}. This Horndeski  hairy black hole soon attracts plenty of attention, and various theoretical and observational phenomena have been disclosed, for instances, thermodynamic and weak gravitational lensing \cite{Walia:2021emv,Atamurotov:2022slw}, the strong gravitational lensing \cite{Kumar:2021cyl}, shadow constraint from EHT observation \cite{Afrin:2021wlj,Vagnozzi:2022moj},  superradiant properties in its rotating counterpart \cite{Jha:2022tdl,Lei:2023wlt},  photon rings in the black hole image \cite{Wang:2023vcv,Hu:2023pyd} and precessing and periodic orbits \cite{Lin:2023rmo}.  These studies disclosed the novel properties brought by the Horndeski hair, differentiated from the Schwarzschild black hole in GR, which are helpful for understanding the gravitational structure of the specific theory.

Another  significant question concerning this Horndeski hairy black hole is its stability.  It was addressed in \cite{Walia:2021emv} that the Horndeski hairy black hole is thermodynamically preferred. Later, we also demonstrated its dynamical stability under various external massless perturbed fields including the scalar field, electromagnetic field and Dirac field in \cite{Yang:2023lcm}.  However, up to now, its dynamical stability has not been completely  investigated.  To promote this topic,  we have to perturb the metric itself and study the linear mode as a first important step, in which the modes of a spherically symmetric black
holes usually can be classified into the even-parity (polar) modes and odd-parity  (axial) modes.

On the other hand, considering that the gravitational waves radiated during black hole coalescence encode rich information about the nature of spacetime, 
it is significant to study the gravitational waves signatures of such hairy black hole.   In particular, the ringdown wave, as the associated gravitational emission at the late time during binary black hole coalescence, is expected to be modeled by the QNM frequencies. At this stage, the QNMs could be the key observable in constructing the patterns of such signal and therefore could be used to examine the no-hair theorem due to its highly dependence on the black hole parameter. 
Thus, the first step in this direction is to study their QNMs, which describe the ringdown stage  after the 
{binary} black holes merger \cite{Berti:2015itd}.  The QNMs are the eigenmodes of a dissipative system, which inherently reflect the effects of all the system's parameters. Consequently, if a black hole acquires an additional Horndeski hair, its QNMs will naturally encode the characteristics of this scalar hair.  Due to the dissipative nature of the system,  the frequencies of QNMs are typically composed of two parts: the real part and the imaginary part, the former  gives the frequency of vibration while the latter describes the damping timescale. Readers can refer to the  reviews \cite{Nollert:1999ji,Berti:2009kk,Konoplya:2011qq} for more details.

Thus, motivated by the above aspects, in this paper we will consider the dynamics of the gravitational odd-parity perturbations of a specific Horndeski hairy black hole. We expect that it has a simpler spectrum than the even-parity modes because it decouples from the perturbations of the scalar field, in contrast to the even-parity modes. We shall study the time-domain evolution and  QNM frequencies for the  odd-parity  perturbations, then further answer the question whether the hairy  black hole is stable under such perturbations.  Moreover, in order to check the imprint of Horndeski hair on the GW in ringdown stage and give some insight into the test of no-hair theorem,  we will use the yielding QNM data to construct the ringdown GW waveform and further calculate the Fisher information matrix (FIM) \cite{Berti:2005ys} to obtain the measurement error of the detection of the Horndeski hair.

This paper is organized as follows. In section \ref{sec:background}, we  derive the modified Regge-Wheeler equation of of the gravitational odd perturbation, and give a proposal to refine it in a specific Horndeski hairy black hole. In section \ref{sec:QNM & error}, we numerically solve the master equations, and extract the QNM frequencies of the gravitational odd perturbation. We analyze the effects of scalar hair $Q$, the angular number $\ell$, and the overtones $n$ on the QNM frequencies. In section \ref{sec:error}, we first briefly review the waveform modeling in the ringdown stage,  and then investigate the detectability of the specific Horndeski hair by calculating the error bound in cases with single-mode and mixed-mode estimations. Section \ref{sec:conclustion} concludes with our key findings and outlooks for future work. Throughout the paper, we will set $c=G=1$ unless we restate it.

\section{Equation of motion for the odd-parity perturbation}\label{sec:background}
\subsection{Review on the action and Horndeski hairy black hole}\label{sec:action}
Horndeski gravity is the most general scalar-tensor field theory in four dimensions which possesses second-order field equations \cite{Horndeski:1974wa}. In this subsection, we shall focus on a special subclass of Horndeski theory and review the static hairy black hole constructed in  \cite{Bergliaffa:2021diw}. The action is
\begin{eqnarray}\label{eq:action}
S=\int d^4x \sqrt{-g}\big[Q_2+Q_3\Box\phi+Q_4R+Q_{4,\chi}\left((\Box\phi)^2-(\nabla^\mu\nabla^\nu\phi)(\nabla_\mu\nabla_\nu\phi)\right)\big],
\end{eqnarray}
which is obtained by setting $Q_5=0$ in the general action \eqref{eq:action0}.
From the above action, we can vary it with respect to $g^{\mu\nu}$ to obtain the field equation
\begin{eqnarray}
Q_4 G_{\mu\nu}=&&\frac{1}{2}(Q_2,_\chi \phi ,_\mu \phi ,_\nu+Q_2 g_{\mu\nu})
+\frac{1}{2}Q_3,_\chi(\phi ,_\mu \phi ,_\nu\square\phi
-g_{\mu\nu} \chi,_\alpha \phi^{, \alpha}+\chi,_\mu \phi ,_\nu
+\chi,_\nu \phi ,_\mu)
\nonumber\\
&&- Q_4,_\chi\Big\{\frac{g_{\mu\nu}}{2}[(\square\phi)^2
-(\nabla_\alpha\nabla_\beta\phi)(\nabla^\alpha\nabla^\beta\phi)-2R_{\sigma\gamma}\phi^{,\sigma}\phi^{,\gamma}]
-\nabla_\mu\nabla_\nu \phi \square\phi
+\nabla_\gamma\nabla_\mu \phi \nabla^\gamma \nabla_\nu \phi
\nonumber\\
&&
-\frac{R}{2}\phi ,_\mu \phi ,_\nu 
+R_{\sigma\mu}\phi^{,\sigma}\phi,_{\nu}+R_{\sigma\nu}\phi^{,\sigma}\phi,_{\mu}+R_{\sigma\nu\gamma\mu} \phi^{,\sigma}\phi^{,\gamma}
\Big\}~~~~~~~~~~~~~~
\nonumber\\
&&-Q_4,_\chi,_\chi \Big\{g_{\mu\nu}(\chi,_{\alpha}\phi^{,\alpha}\square \phi+\chi_{,\alpha} \chi^{, \alpha})+\frac{1}{2}\phi ,_\mu \phi ,_\nu\times
(\nabla_\alpha\nabla_\beta\phi\nabla^\alpha\nabla^\beta\phi-(\square\phi)^2)
- \chi,_\mu \chi,_\nu
\nonumber\\
&&- \square\phi( \chi,_\mu \phi,_\nu
+ \chi,_\nu \phi,_\mu)
- \chi,_\gamma[\phi^{,\gamma}\nabla_\mu\nabla_\nu\phi-(\nabla^\gamma\nabla_\mu\phi)\phi,_{\nu}
-(\nabla^\gamma\nabla_\nu\phi)\phi,_{\mu}]
  \Big\},
\end{eqnarray}
and also obtain the four-current  as 
\begin{eqnarray}\label{eq:current}
j^\nu&=&\frac{1}{\sqrt{-g}}\frac{\delta S}{\delta(\phi_{,\mu})}\nonumber\\
&=&-Q_2,_\chi \phi^{, \nu}-Q_3,_\chi (\phi^{, \nu}\square\phi+\chi^{, \nu})
-Q_4,_\chi (\phi^{, \nu}R-2R^{\nu\sigma}\phi,_\sigma)
\nonumber\\
&&-Q_4,_\chi,_\chi\{\phi^{, \nu}[(\square \phi)^2
-(\nabla_\alpha\nabla_\beta\phi)(\nabla^\alpha\nabla^\beta\phi)]
+2(\chi^{, \nu}\square\phi-\chi,_\mu\nabla^{\mu}\nabla^{\nu}\phi)
\}.
\end{eqnarray}

In order to construct an exact solution to the above sector, we first consider the ansatz
\begin{eqnarray}
\phi&=&\phi(r),\\
ds^2&=&-A(r)dt^2+\frac{dr^2}{B(r)}+r^2(d\theta^2+\sin^2\theta d\varphi^2),
\end{eqnarray}
such that the only non-vanishing component of the current takes the form
\begin{eqnarray}
j^r=-Q_2,_\chi B\phi^{\prime}-Q_3,_\chi\frac{(4A+r A^\prime)}{2rA}B^2{\phi^\prime}^2
+2Q_4,_\chi\frac{B}{r^2 A}[(B-1)A+r B A^\prime]\phi^\prime
- 2Q_4,_\chi,_\chi \frac{B^3(A+r A^\prime)}{r^2A}{\phi^\prime}^3,
\end{eqnarray}
where the prime means the derivative with respect to the coordinate $r$. Then, one can further fix
\begin{eqnarray}\label{Horndeski sol. 1}
Q_2=\alpha_{21} \chi + \alpha_{22}{(-\chi)}^{\omega_2},~~
Q_3=\alpha_{31}{(-\chi)}^{\omega_3},~~
Q_4=\frac{1}{8\pi} + \alpha_{42}{(-\chi)}^{\omega_4},
\end{eqnarray}
and set $\alpha_{21}=\alpha_{31}=0,~~\omega_2=\frac{3}{2}$ and $\omega_4=\frac{1}{2}$ to obtain a solution with  non-trivial scalar field.
Subsequently, imposing the condition {$\displaystyle{\lim_{r \to \infty}}j^r=0$ } yields 
\begin{eqnarray}\label{Horndeski sol. 2}
\phi^{\prime}=\pm\frac{2}{r}\sqrt\frac{{-\alpha_{42}}}{{3B\alpha_{22}}},
\end{eqnarray}
and then solving the field equation gives a non-stealth black hole solution 
\begin{eqnarray}\label{Horndeski sol. 3}
A(r)=B(r)=1 -\frac{2M}{r} + \frac{Q}{r}\ln\left(\frac{r}{2M}\right)~~\text{and}~~Q=8\pi\Big(\frac{2}{3}\Big)^{3/2} \alpha_{42}\sqrt{-\frac{\alpha_{42}}{\alpha_{22}}},
\end{eqnarray}
where $M$ is the integral constant related to the black hole mass and $Q$ is fixed by the parameters of the theory. This indicates that the scalar hair should be classified as a secondary hair rather than a primary scalar hair, the latter requires $Q$ to be an integration constant appearing in the metric \cite{Bakopoulos:2023fmv}. It is noted that $\alpha_{42}$ and $\alpha_{22}$ should have opposite signs in \eqref{Horndeski sol. 3}  to admit a real scalar field. 

Thus, the subclass Horndeski theory with the action \eqref{eq:action} with some specific settings admits a Horndeski hairy black hole with the metric \cite{Bergliaffa:2021diw}
\begin{eqnarray}\label{eq:metric}
ds^2=-f(r)dt^2+\frac{dr^2}{f(r)}+r^2(d\theta^2+\sin^2\theta d\varphi^2)
~~~\mathrm{with}~~~f(r)=1-\frac{2M}{r}+\frac{Q}{r}\ln\left(\frac{r}{2M}\right),
\end{eqnarray}
in which the hairy term appears as a $\log$ term deforming the standard Schwarzschild black hole. $M$ and $Q$ are related to the mass and hairy charge of the black hole. It is obvious that the metric is asymptotically flat since $f(r)\mid_{r\to \infty}=1$.
Let us briefly analyze the structure of the hairy metric \eqref{eq:metric}. For arbitrary $Q$, $r=0$ represents an intrinsic singularity as the curvature scalar is singular, and $f(r)=0$ always admits the solution $r_+=2M$.  In addition, (i) when $Q\to 0$, the metric reduces to the Schwarzschild case. (ii) When $Q>0$, $r_+=2M$ is the unique root of $f(r)=0$, so it has a unique horizon, i.e., the event horizon for the hairy black hole which resembles Schwarzschild black hole. (iii) For $-2M<Q<0$, $f(r)=0$ has two distinguishable positive real roots: the larger one is $r_+=2M$ denoting the position of event horizon, and the smaller one $r_-$ is the Cauchy horizon, lying inside the event horizon. The Cauchy horizon $r_-$ increases as $Q$ decreases, and finally coincides with $r_+$ when $Q=-2M$, indicating the extreme case. 

\subsection{Odd-parity perturbation and modified Regge-Wheeler equation}\label{sec:modified RW}
In this subsection, we will consider the gravitational perturbation to the Horndeski hairy black hole \eqref{eq:metric}. We will focus on the odd-parity perturbation at linear order, thus, we can ignore the coupling between the even-parity  perturbation and the perturbation of the energy-momentum tensor from the scalar field,  both of which decouple from the odd-parity perturbation. The gravitational perturbation of general spherically symmetric black holes in Horndeski gravity was well addressed in \cite{Kobayashi:2012kh,Kobayashi:2014wsa}, based on which we shall review the derivation of the modified Regge-Wheeler equation and present our refined version of this equation. 

Under the Regge-Wheeler gauge \cite{Regge:1957td}, the metric perturbed by the odd-parity mode takes the form
\begin{eqnarray}\label{eq:pertur}
g_{\mu\nu}&= g^{background}_{\mu\nu}+ h_{\mu\nu,\ell m}^{\text{odd}}\qquad\qquad\qquad\quad\nonumber \\
&=\begin{pmatrix}
 A(r)&0&0&0\\
 0&B(r)&0&0\\
 0&0&C(r)&0\\
0&0&0&C(r)\sin^2 {\theta}
 \end{pmatrix}&+
 \begin{pmatrix}
 0&0&0&h_0(r,t)\\
 0&0&0&h_1(r,t)\\
 0&0&0&0\\
h_0(r,t)&h_1(r,t)&0&0
 \end{pmatrix}\sin {\theta} \partial_{\theta}Y_{\ell m}
\end{eqnarray}
where the spherical harmonics can be replaced by Legendre
polynomials by setting the  azmuthal  number $m=0$ without loss of generality, i.e. $Y_{\ell m}\mid_{m=0}=\sqrt{\frac{2\ell+1}{4\pi}}P_\ell(\cos\theta)$, because the background hairy metric is spherically symmetric. By substituting the perturbed metric \eqref{eq:pertur} into the action \eqref{eq:action}, we can obtain a second order on-shell action as
\begin{eqnarray}\label{eq:actionOrder2-1}
S^{(2)}=\int dt dr \left[a_1h_0^2+a_2h_1^2+a_3\left(\dot{h}_1^2 {+}h_0^{\prime2}-2\dot{h}_1h_0^\prime+2\frac{C'}{C}\dot{h}_1h_0\right)\right],
\end{eqnarray}
in which we have integrated the angular distributions with respect to $\theta$ and $\phi$. Here the dot and prime denote the derivative with respect to $t$ and $r$, respectively, and the coefficients are expressed as
\begin{eqnarray}
    a_1&=&\frac{\ell(\ell+1)}{4C}\left[\partial_r \left(C'\sqrt{\frac{B}{A}}{\cal{H}}\right)+\frac{(\ell-1)(\ell+2)}{\sqrt{AB}}\cal{F}\right],\\
a_2&=&-\frac{\ell(\ell+1)(\ell-1)(\ell+2)\sqrt{AB}}{2C}\cal{G},\\
a_3&=&\frac{\ell(\ell+1)}{4} \sqrt{\frac{B}{A}} {\cal{H}},\quad \text{with}\\
{\cal F}&=&2\left(
Q_4+\frac{1}{2}B\phi'\chi'Q_5,_\chi-\chi Q_5,_\phi
\right)\,, \label{def-f}
\\
{\cal Q}&=&2\left[ Q_4-2\chi Q_4,_\chi
+\chi \left(\frac{A'}{2A}B\phi'Q_5,_\chi+Q_5,_\phi\right)
\right]\,, \label{def-g}
\\
{\cal H}&=&2\left[Q_4-2\chi  Q_4,_\chi
+\chi \left(\frac{  C' }{2 C} B \phi' Q_5,_\chi+ Q_5,_\phi\right)\right]\,. \label{def-h}
\end{eqnarray}

To proceed, we introduce an auxiliary field and tortoise coordinate
\begin{align}
\label{eq:auxiliary field}
\Psi(r,t) &= \sqrt{\frac{B C \mathcal{H}^2 Z}{A \mathcal{F}}} \left( \dot{h}_1 - h'_0 + \frac{C'h_0}{C} \right), \\
\label{eq:tortoise coord}
\mathrm{d} r_{*} &= \sqrt{\frac{Z^2}{AB}} \, \mathrm{d}r.
\end{align}

It is noted that comparing against the formulas utilized in \cite{DeFelice:2011ka,Ganguly:2017ort}. Here we introduce an auxiliary function $Z$, whose form can be properly chosen to avoid the superluminal propagation of GWs, as we will see soon. After combining the transformations \eqref{eq:auxiliary field} and \eqref{eq:tortoise coord}, the second order action \eqref{eq:actionOrder2-1} can be reduced as 
\begin{eqnarray}\label{eq:actionOrder2-2}
S^{(2)}=\frac{\ell(\ell+1)}{4(\ell-1)(\ell+2)}\int \mathrm{d}t \mathrm{d}r_{*} \left[\frac{\mathcal{F}}{\mathcal{G}Z^2}\left(\frac{\mathrm{d}\Psi}{\mathrm{d}t}\right)^2-\left(\frac{\mathrm{d}\Psi}{\mathrm{d}r_*}\right)^2-V(r)\Psi^2\right],
\end{eqnarray}
where the effective potential $V(r)$ takes the form
\begin{eqnarray}
V(r)=\frac{(\ell-1) (\ell+2) A \mathcal{F}}{C \mathcal{H}Z^2}-\frac{C^2 }{4 C' Z^2}\left(\frac{A B C'^2}{C^3}\right)'-\frac{C^2 \mathcal{F}^2 }{4\mathcal{F}'}\left(\frac{A B \mathcal{F}'^2}{C^2  \mathcal{F}^3 Z^2}\right)'+\frac{\mathcal{F}^2}{Z'}\left(\frac{A B Z'^2}{4  \mathcal{F}^2 Z^3} \right)'. 
\end{eqnarray}
It is obvious that such second order action is only well-defined for $\ell\ge2$ but not for the modes $\ell=0$ and $\ell=1$ which are in fact not dynamical. $\ell=0$ is spherically symmetric and therefore obeys the Birkhoff theorem such that the perturbation solution is trivial. 
The second order action \eqref{eq:actionOrder2-2}  is ill-defined for $\ell=1$ mode due to the overall term $(\ell-1)^{-1}$, so it should be specially treated. Fortunately, it was found that in GR $\ell=1$ mode  corresponds to an infinitesimal shift of the position of the black hole \cite{Regge:1957td}, while in general scalar-tensor theory $\ell=1$ mode could also correspond to a certain physical perturbation \cite{Kobayashi:2012kh}, which will not  be elaborately interpreted here  because we will focus on the modes $\ell\ge2$ in the current work. 

By varying the action  \eqref{eq:actionOrder2-2}, we obtain the  equation of motion 
\begin{eqnarray}\label{eq:RWeq1}
\left( \frac{\mathrm{d}}{\mathrm{d}r_*^2}-\frac{\mathcal{F}}{\mathcal{G}Z^2}\frac{\mathrm{d}}{\mathrm{d}t^2}-V(r)\right)\Psi=0,
\end{eqnarray}
which is the modified Regge-Wheeler equation for the black hole in  the subclass of Horndeski theory with the action \eqref{eq:action}. 
The radial propagation speed of the odd perturbation can be defined by the coefficient of $\frac{\mathrm{d}}{\mathrm{d}t^2}$ as 
\begin{eqnarray}
c^2_r=\frac{\mathcal{G} Z^2}{\mathcal{F}}.
\end{eqnarray}

\begin{itemize}
  \item \textbf{$Z=1$} 

For $Z=1$, the previous derivation recovers those in \cite{DeFelice:2011ka,Ganguly:2017ort}. Especially, 
the radial propagation speed of the odd perturbation in the current Horndeski hairy background \eqref{eq:metric} is
\begin{eqnarray}\label{eq:oddspeed}
c_r^2=\frac{\mathcal{G}}{\mathcal{F}}=\frac{Q_4-2\chi Q_{4,\chi}}{Q_4 }=\left(1+\frac{3Q}{2r}\right)^{-1}.
\end{eqnarray}
We can obviously see that a non-vanishing scalar charge $Q$ indeed causes the deviation of  $c_r^2$  from the unity, i.e. the vacuum light speed in GR. Thus, the varying of parameter $Q$ may cause $c_r^2<0$, and Laplacian instability will occur. In order to avoid such instability, a condition for keeping a positive $c_r^2$ must be imposed at least at the event horizon, i.e.
\begin{eqnarray}
    Q> - \frac{2}{3}r_+,
\end{eqnarray}
which is consistent with the stability condition addressed in \cite{Kobayashi:2012kh,Ganguly:2017ort}. Moreover, to illustrate how the hair parameter affects the radial propagation speed of the odd perturbation in this case, we plot  FIG.\ref{fig:oddspeed} which shows that the propagation speed can be larger or smaller than the vacuum light speed depending on the hairy charge.
\begin{figure*}[htbp]
\includegraphics[height=7.9cm]{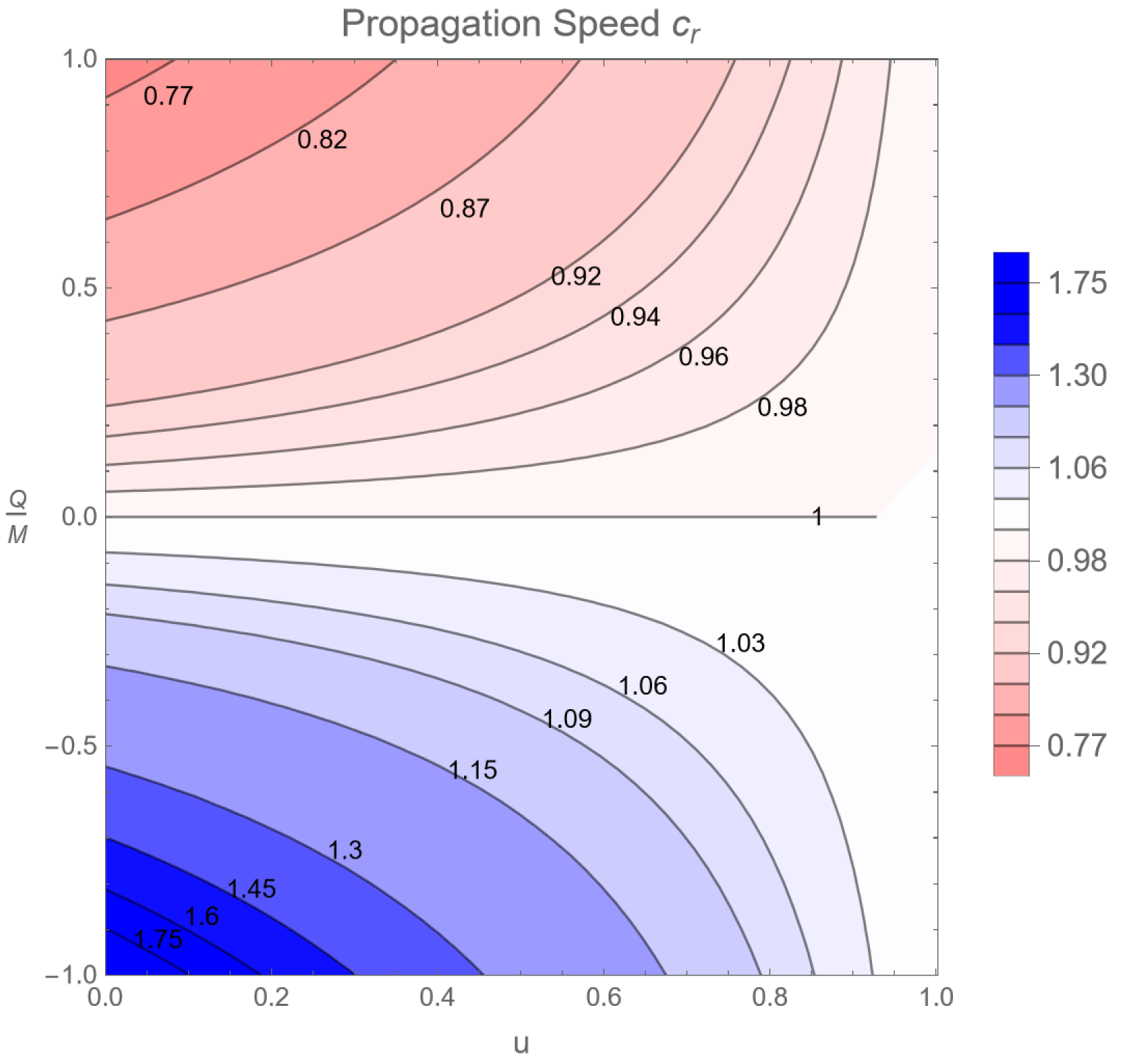}
\caption{\label{fig:oddspeed}The radial propagation speed of odd perturbation,  here we have fixed $M=1/2$ and defined $u=1-\frac{r_+}{r}$.}
\end{figure*}

\item \textbf{$Z=\sqrt{\frac{\mathcal{F}}{\mathcal{G}}}=\sqrt{1+\frac{3Q}{2r}}$} 

It is explicit that for $Z=\sqrt{\frac{\mathcal{F}}{\mathcal{G}}}$, the modified Regge-Wheeler equation \eqref{eq:RWeq1} becomes a standard wavelike equation, and the propagation speed $c_r$ returns to unity, same as that in GR. Our analysis indicates that with proper auxiliary field and tortoise coordinate, the superluminal propagation in previous study \cite{Ganguly:2017ort} can be mathematically cured.
\end{itemize}

Solving the equation \eqref{eq:RWeq1} with proper boundary conditions allows us to extract the spectrum of the odd-parity perturbation and to further analyze the stability of the hairy black hole and other related properties. In principle, the transformations \eqref{eq:auxiliary field} and \eqref{eq:tortoise coord} are expected not to affect the dynamic of the perturbation. In order to check this, we shall solve the equations \eqref{eq:RWeq1} for both $Z=1$ and $Z=\sqrt{1+\frac{3Q}{2r}}$ to extract the QNM spectrum, yielding the same results as shown in the next section.

\section{Quasinormal mode}\label{sec:QNM & error}
Recalling the Horndeski hairy black hole \eqref{eq:metric}, the modified Regge-Wheeler equation \eqref{eq:RWeq1} can be rewritten as 

\begin{eqnarray}\label{eq:RWeq2}
&&\left[\frac{\mathrm{d}}{\mathrm{d}\bar{r}_*^2}-\left(1+\frac{3Q}{2r}\right)\frac{\mathrm{d}}{\mathrm{d}t^2}-V_1(r)\right]\Psi_{1}=0~~~~\text{with}~~~~
\mathrm{d}\bar{r}_{*}=f^{-1}\mathrm{d}r,\\
&&V_1(r)=\frac{f^2 \left(27 Q^2+48 Q r+32 r^2\right)}{4 r^2 (3 Q+2 r)^2}+\left(1+\frac{3Q}{2r}\right)f\left[\frac{(\ell+2)(\ell-1)}{r^2}-\frac{ f' (3 Q+4 r) }{ (3 Q+2 r)^2}\right]
\end{eqnarray}
for $Z=1$, and
\begin{eqnarray}\label{eq:RWeq3}
&&\left[\frac{\mathrm{d}}{\mathrm{d}\tilde{r}_*^2}-\frac{\mathrm{d}}{\mathrm{d}t^2}-V_2(r)\right]\Psi_{2}=0~~~~\text{with}~~~~
\mathrm{d}\tilde{r}_{*}=f^{-1}\left(1+\frac{3Q}{2r}\right)\mathrm{d}r,\\
&&V_2(r)=\frac{f^2 \left(135 Q^2+240 Q r+128 r^2\right)}{8 r (3 Q+2 r)^3}+f \left[\frac{(\ell+2)(\ell-1)}{r^2}-\frac{f' (9 Q+8 r)}{2 (3 Q+2 r)^2}\right]
\end{eqnarray}
for $Z=\sqrt{\frac{\mathcal{F}}{\mathcal{G}}}=\sqrt{1+\frac{3Q}{2r}}$.
Both of them revert to the standard Regge-Wheeler potential \cite{Regge:1957td} when the Horndeski hair vanishes. Since the explicit expression for the master equations are clearly established,  we are now in a position to solve the equation and extract the dynamical properties for the odd-parity perturbation, which allows us to work in both the time and frequency domains. 

\subsection{Preparation}
Firstly, we will outline the method and boundary conditions in our numerical calculations to solve \eqref{eq:RWeq2} in the time domain, and (\ref{eq:RWeq2},  \ref{eq:RWeq3}) in the frequency domain.

\begin{itemize}
    \item  \textbf{In the time domain}, we shall work in tortoise coordinate grids and apply the finite difference method (Readers may refer to \cite{Zhu:2014sya, Yang:2023lcm} for more details) to discretize the radial coordinate $r=r(\bar{r}_*)=r(j\Delta \bar{r}_{*})=r_j$, thus the perturbation and the potential are discretized as $\Psi(t,r)=\Psi(i\Delta t,j\Delta \bar{r}_{*})=\psi_{i,j}$ and $V(r)=V(j \Delta \bar{r}_{*})=V_j$. Then we use the central difference to discretize the derivative of the perturbation and rewrite the equation \eqref{eq:RWeq2} into an iterative equation
\begin{eqnarray}
    \Psi _{i+1,j}=\frac{\Delta t^2 \left(\Psi _{i,j-1}-2 \Psi _{i,j}+\Psi _{i,j+1}\right)}{\left(1+\frac{3Q}{2r_j}\right)\Delta \bar{r}_{*}^2}-\Delta t^2 V_j \Psi _{i,j}+2 \Psi _{i,j}-\Psi _{i-1,j}.
\end{eqnarray}
To ensure reliable numerical results in solving partial differential equation, the Von Neumann stability condition, also known as Fourier stability analysis, should be analyzed to ensure that errors introduced during the computation do not grow uncontrollably as the calculation progresses. In this scenario, the Von Neumann stability condition usually requires $\frac{\Delta t}{\Delta \bar{r}_{*}}\leq 1$  which can make the numerical scheme stable \cite{Smith1985}, such that we set the intervals $\Delta \bar{r}_{*}=2\Delta t=0.1$. Moreover, we will choose the Gaussian packet $\Psi_{i<0,j}=0$,  $\Psi_{0,j}=\exp[-\frac{(r_{j}-18)^2}{2}]$ as the initial data to evolve the iterative equation. The evolutionary profile of the perturbation typically includes an initial outburst, a damped oscillation, and a late-time tail, from which one can compare the damping time of the perturbation in varying black hole parameters and roughly predict the (in)stability of the background spacetime.

\item \textbf{In the frequency domain}, we shall factor out the time dependency of  the perturbations by separating $\Psi_{I}(t,r)=e^{-i\omega t}R_{I}(r)$ with $I=1,2$ denoting the fields in  \eqref{eq:RWeq2} and \eqref{eq:RWeq3},  which yield general eigenvalue equations
\begin{align}
\label{eq:master eq1}
    \left[\frac{\mathrm{d}}{\mathrm{d}\bar{r}_*^2}+\left(1+\frac{3Q}{2r}\right)\omega^2-V_{1}(r)\right]R_{1}(r)=0, ~~~\text{for}~~Z=1,\\
  \label{eq:master eq2}
    \left[\frac{\mathrm{d}}{\mathrm{d}\tilde{r}_*^2}+\omega^2-V_{2}(r)\right]R_{2}(r)=0, ~~~\text{for}~~Z=\sqrt{1+\frac{3Q}{2r}}.
\end{align}
Their corresponding boundary conditions for the QNMs solution are ingoing at the horizon
\begin{eqnarray}
  \label{eq:horizonBdy1} R_{1}(r\to r_+)&\sim& e^{-i\sqrt{1+\frac{3Q}{2r_+}}\omega \bar{r}_*}, \\
  \label{eq:horizonBdy2} R_{2}(r\to r_+)&\sim& e^{-i\omega \tilde{r}_*},
\end{eqnarray}
and outgoing at the infinity
\begin{eqnarray}
  \label{eq:infinityBdy1} R_{1}(r\to \infty)\sim e^{i\omega \bar{r}_*},\\
   \label{eq:infinityBdy2}  R_{2}(r\to \infty)\sim e^{i\omega \tilde{r}_*},
\end{eqnarray}
respectively. Thus, once the black hole parameters (for instance, the mass $M$ and the hairy charge $Q$) are fixed, the eigenvalues could be solved as an infinite series of complex numbers, which are the so-called QNM frequencies. Similar to the normal modes, the real part of the QNMs represents the physical oscillation angular frequency, but they differ from the normal modes by having a non-vanishing imaginary part that represents the exponential damping of the perturbation in a dissipating system. 

We shall employ the pseudospectral method and the matrix method to solve the master equation \eqref{eq:master eq1} with the bondary conditions \eqref{eq:horizonBdy1} and \eqref{eq:infinityBdy1}, to firstly show the validity of the obtained QNMs. Then we apply the pseudospectral method to solve \eqref{eq:master eq2} with the boundary conditions \eqref{eq:horizonBdy2} and \eqref{eq:infinityBdy2}, to verify the isospectrality between the two choices of $Z$. Since those algorithms are widely used in studies of black hole spectroscopy, here we shall skip the main steps, and readers may refer to \cite{Jansen:2017oag} for the pseudospectral method and \cite{Lin:2019mmf} for the matrix method respectively.  It is noted that due to the $\log$ term in the metric function, a redefinition of the radial wave function is crucial when using  both methods, as we have addressed in appendix B of our previous work \cite{Yang:2023lcm}. Similar processes will be performed in the current numerical calculations.

\end{itemize}

Following the above preparation, we shall show our numerical results of the eigenfrequencies of the QNMs and time evolution of perturbed field in next subsection. In our numeric, we find that as the hair parameter goes closer to the extremal case, the precision of our calculation becomes worse and even the code may break down. So, we shall focus on the parameter regime $-M\leq Q\leq M$, and for simplicity, we also rescale the Horndeski charge into the dimensionless $\tilde{Q}=Q/M$ in the calculations.

\subsection{Numerical results}\label{sec:QNM results}
The QNM frequencies with nodes $n=0,1$ for the selected hairy charges  are listed in TABLE \ref{Table:GWoddQNMgirds},  where we ensure the accuracy of the results by evaluating their relative error. It is shown that for the $Z=1$ case, the pseudospectral and matrix methods achieve high accuracy in the nodes $n=0$ and $1$, and their results are in good agreement with each other, with a relative error on the order of $10^{-5}$ or even less. 
For the $Z=\sqrt{\mathcal{F}/\mathcal{G}}$ case, the QNM results obtained from the pseudospectral method are highly consistent with the results of the $Z=1$ case, with a relative error of less than $10^{-5}$. This demonstrates the isospectrality between the two choices of $Z$. 

\begin{table}[h]
	\centering
	\begin{tabular}{|c|c|c|c|c|c|c|c} \cline{1-7}
		\multicolumn{3}{|c}{}& &  $\tilde{Q}=0$&  $\tilde{Q}=-0.5$&  $\tilde{Q}=-1$ & \\ \cline{1-7}
		\multirow{6}{2.5em}{$\ell=2$}&  \multirow{5}{2.5em}{$n=0$} &\multirow{3}{5em}{$Z=1$}& PS   & 0.747343 - 0.177925 i & 0.679390 - 0.163833 i & 0.617493 - 0.153284 i & \\ \cline{4-7} 
&&&\multirow{2}{1.5em}{M}  & 0.747343 - 0.177925 i & 0.679390 - 0.163833 i & 0.617495 - 0.153284 i & \\ 
&& &  &           $\mathcal{O}(10^{-6})$; $\mathcal{O}(10^{-6})$             &          $\mathcal{O}(10^{-6})$; $\mathcal{O}(10^{-6})$            &    $\mathcal{O}(10^{-6})$; $\mathcal{O}(10^{-6})$  &\\ \cline{3-7}
		  & & \multirow{2}{5em}{$Z=\sqrt{\mathcal{F}/\mathcal{G}}$}&\multirow{2}{1.5em}{PS}  
	 & 0.747343 - 0.177925 i & 0.679390 - 0.163833 i & 0.617493 - 0.153284 i & \\ 
& & &&   $\mathcal{O}(10^{-11})$; $\mathcal{O}(10^{-11})$   &    $\mathcal{O}(10^{-9})$; $\mathcal{O}(10^{-8})$  &   $\mathcal{O}(10^{-8})$; $\mathcal{O}(10^{-7})$&\\ \cline{2-7}
		&  \multirow{5}{2.5em}{$n=1$} &\multirow{3}{5em}{$Z=1$} &
PS  & 0.693422 - 0.547830 i & 0.623772 - 0.504366 i & 0.573985 - 0.481447 i & \\ \cline{4-7} 
& & &\multirow{2}{1.5em}{M}  & 0.693420 - 0.547829 i & 0.623768 - 0.504368 i & 0.574023 - 0.481421 i & \\  
& & &&   $\mathcal{O}(10^{-6})$; $\mathcal{O}(10^{-6})$   &   $\mathcal{O}(10^{-6})$; $\mathcal{O}(10^{-6})$  &   $\mathcal{O}(10^{-5})$; $\mathcal{O}(10^{-5})$   &\\ \cline{3-7}
		&  &  \multirow{2}{5em}{$Z=\sqrt{\mathcal{F}/\mathcal{G}}$}&\multirow{2}{1.5em}{PS} 
	&0.693422 - 0.547830 i & 0.623772 - 0.504366 i & 0.573993 - 0.481448 i & \\  
& && & $\mathcal{O}(10^{-10})$; $\mathcal{O}(10^{-10})$    &  $\mathcal{O}(10^{-6})$; $\mathcal{O}(10^{-8})$  &  $\mathcal{O}(10^{-5})$; $\mathcal{O}(10^{-7})$  &\\ \cline{1-7}
		\multirow{6}{2.5em}{$\ell=3$}&  \multirow{5}{2.5em}{$n=0$} &\multirow{3}{5em}{$Z=1$}&
PS  & 1.198887 - 0.185406 i & 1.078947 - 0.167580 i & 0.955588 - 0.150350 i & \\ \cline{4-7} 
& & &\multirow{2}{1.5em}{M}  & 1.198887 - 0.185406 i & 1.078947 - 0.167580 i & 0.955587 - 0.150349 i & \\  
& &&&               $\mathcal{O}(10^{-6})$; $\mathcal{O}(10^{-6})$           &            $\mathcal{O}(10^{-6})$; $\mathcal{O}(10^{-6})$          &   $\mathcal{O}(10^{-6})$; $\mathcal{O}(10^{-6})$   &\\ \cline{3-7}
	    &  &  \multirow{2}{5em}{$Z=\sqrt{\mathcal{F}/\mathcal{G}}$}&\multirow{2}{1.5em}{PS} 
	  & 1.198887 - 0.185406 i & 1.078947 - 0.167580 i & 0.955588 - 0.150350 i & \\ 
& & &&  $\mathcal{O}(10^{-13})$;$\mathcal{O}(10^{-11})$  &   $\mathcal{O}(10^{-9})$; $\mathcal{O}(10^{-8})$   &   $\mathcal{O}(10^{-8})$; $\mathcal{O}(10^{-8})$ &\\ \cline{2-7}
		&  \multirow{5}{2.5em}{$n=1$} &\multirow{3}{5em}{$Z=1$} &
PS& 1.165288 - 0.562596 i & 1.046267 - 0.508376 i & 0.929446 - 0.458644 i & \\ \cline{4-7} 
& & &\multirow{2}{1.5em}{M}  & 1.165288 - 0.562596 i & 1.046269 - 0.508377 i & 0.929442 - 0.458645 i & \\  
& && &              $\mathcal{O}(10^{-6})$; $\mathcal{O}(10^{-6})$           &   $\mathcal{O}(10^{-6})$; $\mathcal{O}(10^{-6})$  &  $\mathcal{O}(10^{-6})$; $\mathcal{O}(10^{-6})$  &\\ \cline{3-7}
		&  &  \multirow{2}{5em}{$Z=\sqrt{\mathcal{F}/\mathcal{G}}$}&\multirow{2}{1.5em}{PS} 
	&  1.165288 - 0.562596 i & 1.046267 - 0.508376 i & 0.929442 - 0.458641 i & \\  
& &&&  $\mathcal{O}(10^{-11})$; $\mathcal{O}(10^{-11})$    &  $\mathcal{O}(10^{-7})$; $\mathcal{O}(10^{-7})$  &  $\mathcal{O}(10^{-6})$; $\mathcal{O}(10^{-6})$  &\\ \cline{1-7}
		\multirow{6}{2.5em}{$\ell=4$}&  \multirow{5}{2.5em}{$n=0$} &\multirow{3}{5em}{$Z=1$}& 
PS & 1.618357 - 0.188328 i & 1.451873 - 0.169314 i & 1.274651 - 0.149665 i & \\ \cline{4-7} 
& & &\multirow{2}{1.5em}{M}  & 1.618357 - 0.188328 i & 1.451873 - 0.169314 i & 1.274651 - 0.149666 i & \\  
& &&&            $\mathcal{O}(10^{-6})$; $\mathcal{O}(10^{-6})$          &             $\mathcal{O}(10^{-6})$; $\mathcal{O}(10^{-6})$           &   $\mathcal{O}(10^{-6})$; $\mathcal{O}(10^{-6})$ &\\ \cline{3-7}
		&  &  \multirow{2}{5em}{$Z=\sqrt{\mathcal{F}/\mathcal{G}}$}&\multirow{2}{1.5em}{PS} 
	& 1.618357 - 0.188328 i & 1.451873 - 0.169314 i & 1.274651 - 0.149665 i & \\  
& &&&             $\mathcal{O}(10^{-13})$; $\mathcal{O}(10^{-12})$         &   $\mathcal{O}(10^{-9})$; $\mathcal{O}(10^{-9})$   &   $\mathcal{O}(10^{-9})$; $\mathcal{O}(10^{-8})$ &\\ \cline{2-7}
		&  \multirow{5}{2.5em}{$n=1$} &\multirow{3}{5em}{$Z=1$}&
PS& 1.593263 - 0.568669 i & 1.427902 - 0.511200 i & 1.255580 - 0.452830 i & \\ \cline{4-7} 
& & &\multirow{2}{1.5em}{M}  & 1.593263 - 0.568669 i & 1.427904 - 0.511200 i & 1.255583 - 0.452831 i & \\  
& &&&    $\mathcal{O}(10^{-6})$; $\mathcal{O}(10^{-6})$  &   $\mathcal{O}(10^{-6})$; $\mathcal{O}(10^{-6})$    & $\mathcal{O}(10^{-6})$; $\mathcal{O}(10^{-6})$   &\\ \cline{3-7}
		&  &  \multirow{2}{5em}{$Z=\sqrt{\mathcal{F}/\mathcal{G}}$}&\multirow{2}{1.5em}{PS} 
	&  1.593263 - 0.568669 i & 1.427902 - 0.511201 i & 1.255579 - 0.452831 i & \\  
& &&&              $\mathcal{O}(10^{-11})$; $\mathcal{O}(10^{-11})$         &   $\mathcal{O}(10^{-8})$; $\mathcal{O}(10^{-7})$ &  $\mathcal{O}(10^{-7})$; $\mathcal{O}(10^{-6})$   &\\ \cline{1-7}
\end{tabular}
\caption{The QNMs solved from the equations of the odd perturbation \eqref{eq:RWeq1} of the non-rotating Horndeski hairy black hole \eqref{Horndeski sol. 3}, for $\ell=2,3,4$ and $n=0,1$. For the $Z=1$ case, the QNM data are obtained by the pseudospectral (abbreviation as PS) and matrix (abbreviation as M) methods, and their relative errors are evaluated  as $\Delta_k=\mid\frac{\text{M}(\omega_k)-\text{PS}(\omega_k)}{\text{PS}(\omega_k)}\mid$ with $k=\text{Re}, \text{Im}$. For the $Z=\sqrt{\mathcal{F}/\mathcal{G}}$ case, the data are determined by the pseudospectral method, and their errors are defined as the deviation of QNMs frequencies between two choices of $Z$, i.e. $\Delta_k=\mid\frac{\text{PS}(\omega'_k)-\text{PS}(\omega_k)}{\text{PS}(\omega_k)}\mid$. In the cells, we give the corresponding errors for real and imaginary of the frequencies,  repectively. 
        \label{Table:GWoddQNMgirds}}
\end{table}

\begin{figure*}[htbp]
\includegraphics[height=7.9cm]{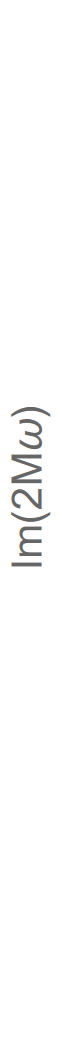}\hspace{0.1cm}
\includegraphics[height=7.9cm]{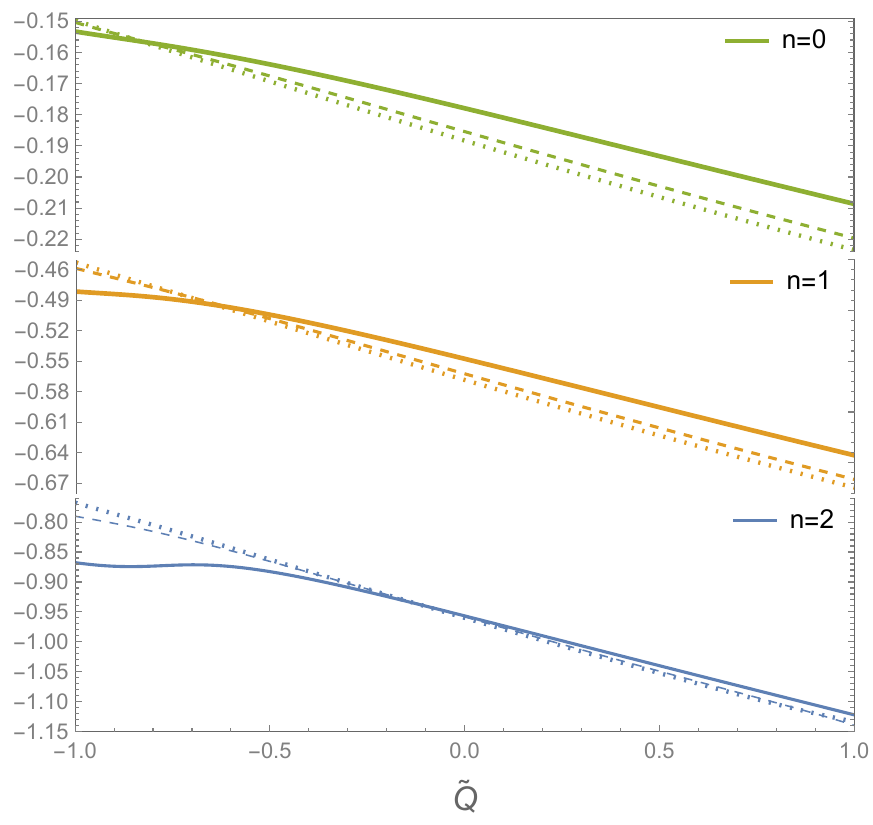}
\includegraphics[height=7.9cm]{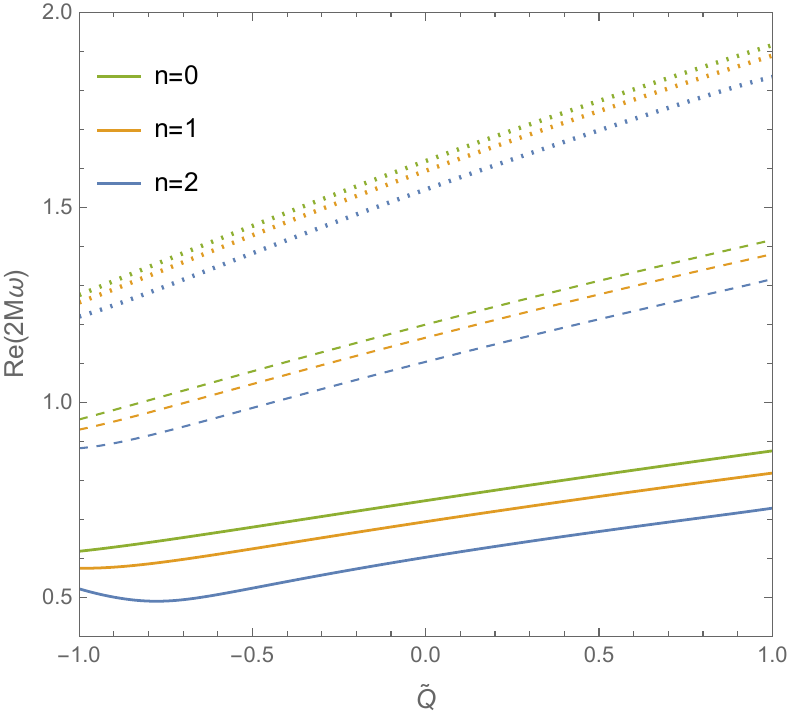}
\caption{\label{GWoddQNF}The dimensionless QNMs as functions of the hair parameter $\tilde{Q}$ for various modes. The different curves are denoted by the angular momenta $\ell=$ 2 (solid),3 (dashed),4 (dotted) and nodes $n=$ 0 (green),1 (orange),2 (blue), respectively.}
\end{figure*}

\begin{figure*}[htbp]
\includegraphics[height=3.8cm]{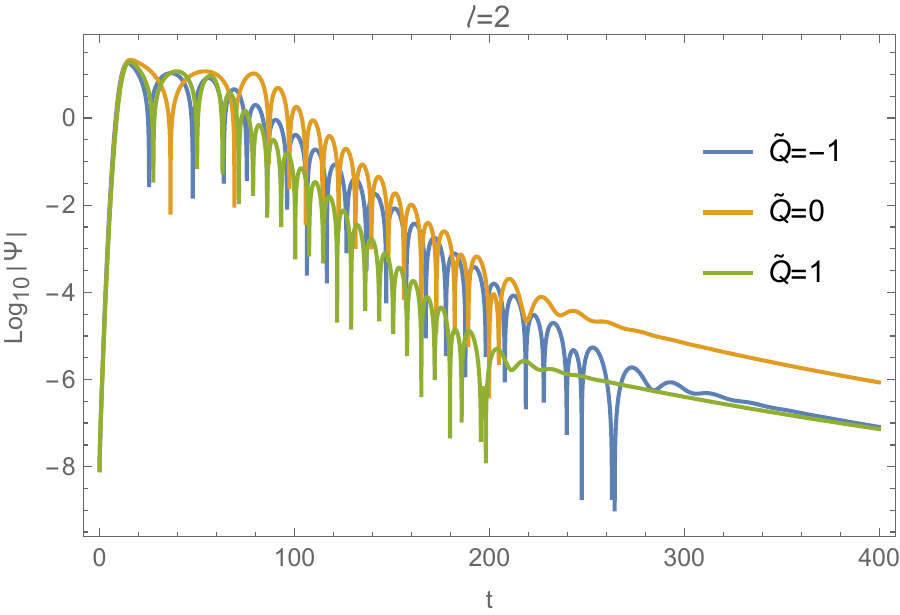}\hspace{0.1cm}
\includegraphics[height=3.8cm]{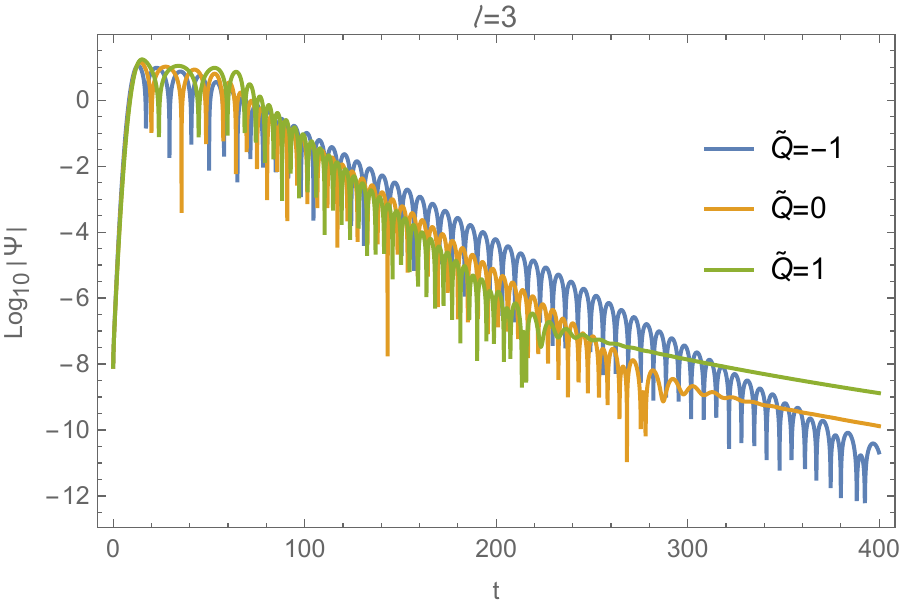}\hspace{0.cm}
\includegraphics[height=3.8cm]{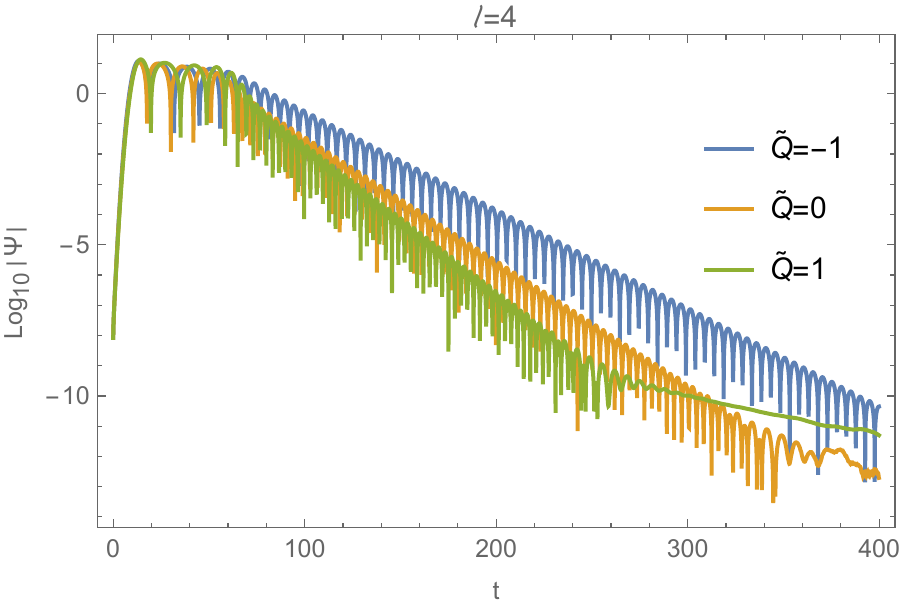}
\caption{\label{GWoddTimedomain}The time evolution of the gravitational odd perturbation of the modes $\ell=2~(\text{left}),3~(\text{middle}),4~(\text{right})$ for selected hairy charge. The lifetime can be estimated by relating the amplitude with decay rate as $\Psi\sim e^{-\alpha \, t}$. In the above $\log_{10}$ plots, the decay rate $\alpha$ can be even intuitively inferred  with the slope of the amplitude $\log_{10}|\Psi|\sim - \alpha \, t$. Therefore, the lifetime of a given mode is inversely proportional to the slope of its amplitude $\tau \sim \frac{1}{\alpha}$ in the ringing stage. From the above plots, we observe that perturbations with a steeper slope have a shorter lifetime, and vice versa. }
\end{figure*}

In order to explicitly see the effect of the Horndeski hair on the QNMs, we plot the frequencies with the first three nodes as functions of hairy charge for $\ell=2,3,4$ modes in FIG.\ref{GWoddQNF}. For all the modes except $\ell=n=2$ (the blue solid curves), the imaginary part increases while the real part decreases as  the scaled hair parameter becomes smaller, but the imaginary part, related to the damping rates by $-Im(\omega)\sim1/\tau$,  is always negative. For one thing, this causes the odd perturbation to always decay, and therefore the black hole can remain stable in the range of $\tilde{Q}\in\{-1,1\}$. For another, this causes the odd perturbation to have a longer damping time. This is consistent with the phenomena we can observe in time domain (FIG.\ref{GWoddTimedomain}), that the ringing stage of the perturbation with $\tilde{Q}=-1$ is always lasting longer than that for other parameters. In addition, for the tendency of the real part $Re(\omega)$, it can be obviously seen that the decreasing  of $\tilde{Q}$ can suppress the oscillation of the odd perturbation, which corresponds to a sparse oscillations for smaller $\tilde{Q}$ in the time domain.
Regarding the effects of angular momentum and overtone, we find that the influence of $\ell$ on $Im(\omega)$ is much weaker than that of $n$. Moreover, it is interesting to notice that as $\tilde{Q}$ becomes sufficiently negative, the long-lived modes with the same node are replaced by the $\ell>2$ modes, instead of the typical $\ell=2$ mode in the usual GR scenario. For higher overtones, the $Im(\omega)$ of $\ell=2$ mode is highly suppressed by the negative hairy charge. In particular,  both $Im(\omega)$ and $Re(\omega)$ for the mode $\ell=n=2$  will slightly change their trend at  $\tilde{Q}\sim-0.8$, indicating novel effects of the hairy charge on the higher overtones. 

\begin{figure*}[htbp]
\includegraphics[height=5cm]{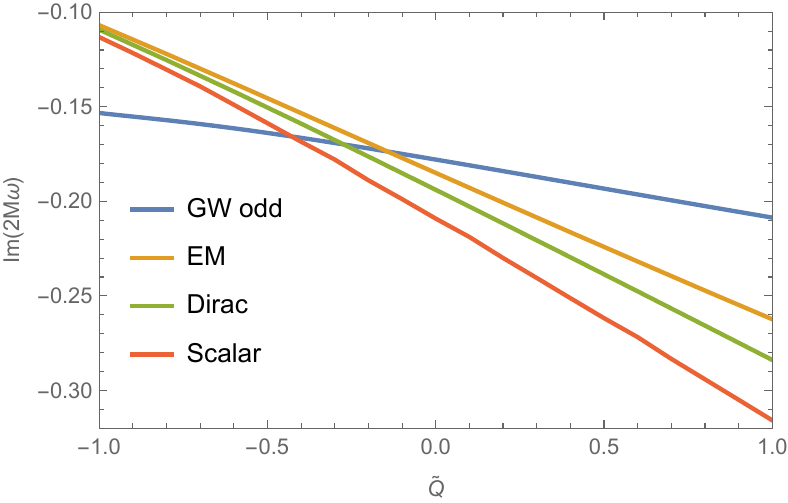}\hspace{0.5cm}
\includegraphics[height=5cm]{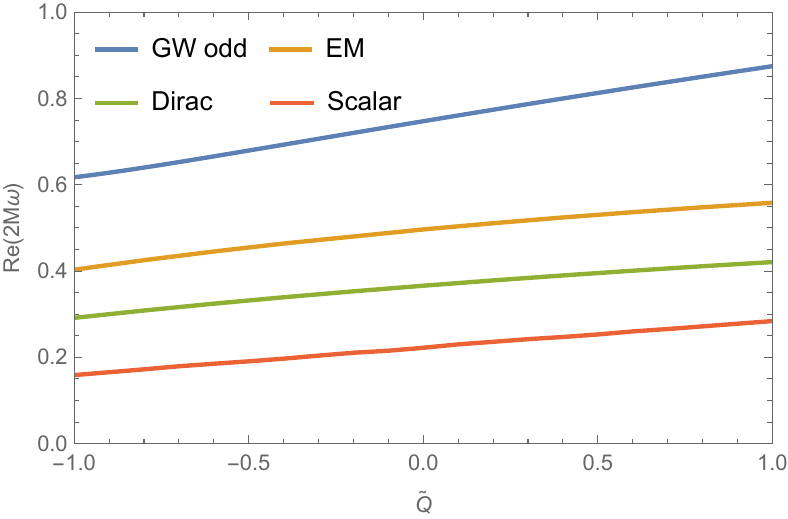}\\
\caption{\label{AllspinQNF}The fundamental 
 QNMs as functions of the hair parameter $\tilde{Q}$, for the perturbed massless scalar 
 ($s=0,\ell=0$), Dirac ($s=1/2,\ell=1$), electromagnetic 
 ($s=1,\ell=1$) and GW odd ($s=2,\ell=2$) fields.}
\end{figure*}

\begin{figure*}[htbp]
\includegraphics[height=4.5cm]{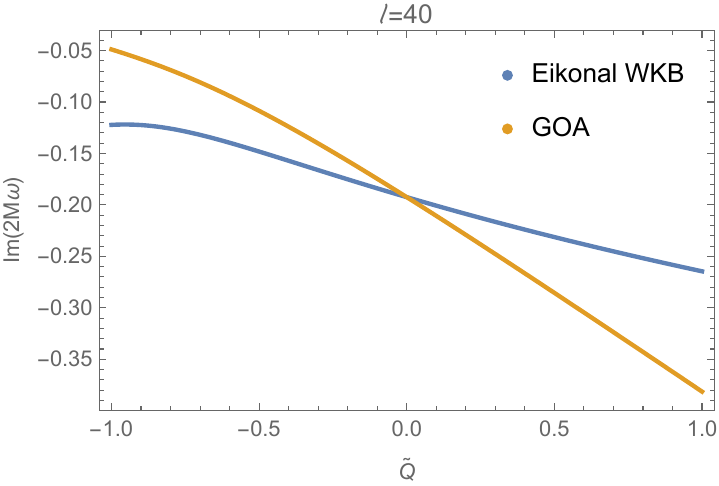}
\includegraphics[height=4.5cm]{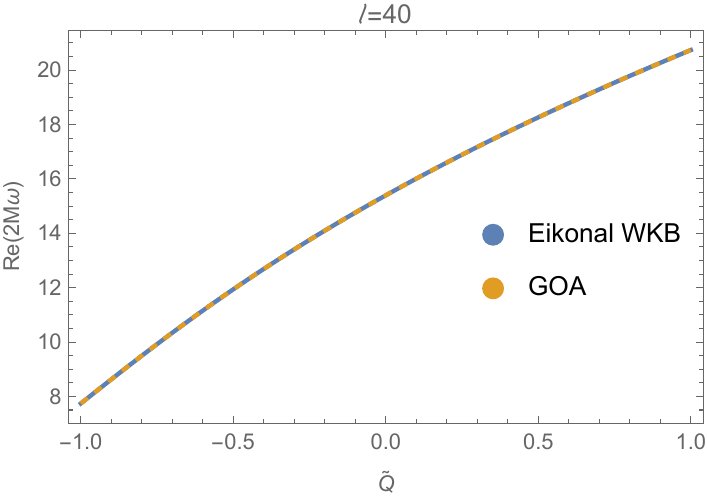}
\caption{\label{QNFlargeL}The fundamental QNMs as functions of the hair parameter $\tilde{Q}$. Here we have fixed $\ell=40$. The blue curve represents the QNM obtained from the exact WKB formula \eqref{eq:eikonal} for the odd perturbation in the large-$\ell$ limit. The orange curve represents the QNM obtained from the geometric-optics approximation formula \eqref{eq:GOA}, which is the QNM for scalar, Dirac and electromagnetic field perturbations degenerated to  in the large-$\ell$ limit. The mismatch of the imaginary part between $\omega_{\text{eikonal}}$ and $\omega_{\text{GOA}}$ obviously arises for nonvanishing Horndeski hair.}
\end{figure*}

In our previous work \cite{Yang:2023lcm}, we addressed the QNMs of massless scalar field, electromagnetic field and Dirac field around the Horndeski hairy black hole, so we collect those fundamental QNMs frequencies and the current QNMs of odd perturbation into FIG.\ref{AllspinQNF}, and shall discuss the effect of hair parameter on how the QNMs depend on the field's spin. Our results show that for positive $\tilde{Q}$, the  odd perturbation has the longest damping time and the strongest oscillation in their fundamental ($n=0$) and lowest-lying modes. However, as $\tilde{Q}$ is approaching $-1$, the odd perturbation turns into the shortest-lived field. 
To further understand the abnormal behavior of the odd perturbation at the analytic level, we shall analyze its dynamics in the eikonal ($\ell>>1$) limit, in which the potential of the standard wave equation \eqref{eq:master eq2} reduces to 
\begin{eqnarray}\label{eq:Ve}
    V_{e}(r)=\ell^2 \frac{f}{r^2},
\end{eqnarray}
and the QNM frequency can be analytically extracted as  \cite{PhysRevD.35.3621}
\begin{eqnarray}\label{eq:eikonal}
    \omega_{\text{eikonal}}=\sqrt{V_e(r_0)}-i(n+\frac{1}{2})\sqrt{\frac{-1}{2V_e(r_0)}\left(\frac{\mathrm{d^2}}{\mathrm{d}\tilde{r}^2_*}V_{e}\right)_{r=r_{0}}} \, ,
\end{eqnarray}
where $r_0$ is the position where $V_{e}(r)$ reaches its maximum \footnote{This process of extracting QNMs frequencies in eikonal limit is not proper for the master equation \eqref{eq:master eq1} because it is not the standard wave equation for WKB analysis.}. It is noted that the effective potential \eqref{eq:Ve} is the same as those for various massless matter field perturbations \cite{Yang:2023lcm} for which in the eikonal limit, their QNM frequencies can also be calculated by the geometric-optics approximation formula  \cite{Cardoso:2008bp}
\begin{eqnarray}\label{eq:GOA}
&\omega_{\text{GOA}}=\ell ~ \Omega_{c}-i(n+\frac{1}{2})|\lambda_{LE}|,\\
&\text{with} \quad \Omega_{c}=\sqrt{\frac{f}{r_{c}}}\quad \text{and} \quad |\lambda_{LE}|=\sqrt{-\frac{r^2_c}{2f_c}\left(\frac{\mathrm{d^2}}{\mathrm{d}\bar{r}^2_{*}} \frac{f}{r^2}\right)_{r=r_{c}}}.
\end{eqnarray}
Here $r_c$ is the radius of the null circular  orbit, given by the positive root of the equation $2f(r) - r f'(r)=0$, so we have $r_c=r_0$. Carefully comparing \eqref{eq:eikonal} and \eqref{eq:GOA}, we find that their real part are consistent, but the imaginary parts mismatch because the tortoise coordinates, $\tilde{r}_{*}$ and  $\bar{r}_{*}$, are different. This implies that though the link between the null geodesics and quasinormal modes for the massless extrenal field perturbations is fulfilled, it is violated for the gravitational odd perturbation in the current theory. Note that  a similar violation was also found in Einstein-Lovelock theory, in which the effective potential of the gravitational perturbation in the eikonal limit was different from radial potential of the null particle \cite{Konoplya:2017wot}. Additionally, we plot FIG. \ref{QNFlargeL}  to show how the Horndeski hair affect the mismatch between $\omega_{\text{eikonal}}$ and $\omega_{\text{GOA}}$ for the odd perturbation.

From the investigation on the QNMs, we find that the effect of the hair parameter apparently differs the BH in Horndeski gravity from GR by significantly influencing the lifetime and the oscillation of gravitational odd perturbation. This will help to recognize and distinguish the hairy black hole in constructing the ringdown wave pattern and even constrain the hair parameter in the ringdown observation. An interesting question comes follow and says that,  based on the yielding QNM data, to what extent can we constrain the horndeski BH hair according to the ringdown signal captured by GW interferometers. We will partly discuss this issue in next section.

\section{Hair parameter estimation by detection of ringdown wave}\label{sec:error}

In this section, we will determine the  measurement errors of the hair parameter $\tilde{Q}$ by the GW detectors, which can be estimated by computing the Fisher information matrix (FIM) \cite{Berti:2005ys}. Considering that the realistic GW events are usually related to the rotating black holes, the  studies based on QNM spectrum from non rotating may not be applied in practice. However, the preliminary analysis on non-rotating scenario is still important to be carried out because it encodes the leading order information for sufficiently slow rotation cases. To this end, we start with a popular template of ringdown signals in GR
\begin{eqnarray}\label{eq:signal}
    h=h_{+}F_{+}+h_{\times}F_{\times},
\end{eqnarray}
where $F_{+,\times}(\theta,\phi)$ are the detector pattern functions that depend on the propagation direction of the wave and the orientation of the detection plane.  The strain functions $h_{+,\times}(t)$ of two polarizations at ringdown stage are given by 
\begin{eqnarray}\label{eq:strain}
 h_{+}(t)&=&\sum_{\ell mn} A^{+}_{\ell mn}e^{-\frac{\pi t f_{\ell m n}}{\mathcal{Q}_{\ell mn}}}S_{\ell m n}\;\cos(\phi^{+}_{\ell mn}+2 \pi t f_{\ell mn}),\\
 h_{\times}(t)&=&\sum_{\ell mn} A^{\times}_{\ell mn}e^{-\frac{\pi t f_{\ell m n}}{\mathcal{Q}_{\ell mn}}}S_{\ell m n}\;\sin(\phi^{\times}_{\ell mn}+2 \pi t f_{\ell mn}),
\end{eqnarray}
where $\phi^{+,\times}_{\ell mn}$ and $A^{+,\times}_{\ell mn}$ represent the phase angles and the reduced amplitudes that have encapsulated any constant normalization factors. $S_{\ell mn}(\theta,\phi)$ denotes the spin-weighted spheroidal harmonics. The quality factors $f_{\ell m n}$ and $\mathcal{Q}_{\ell mn}$ are connected with the QNM frequency via
\begin{eqnarray}
    \omega_{\text{QNM}}=2\pi f_{\ell mn}+\frac{i}{\tau_{\ell mn}}, \quad \mathcal{Q}_{\ell mn}=\pi f_{\ell mn}\tau_{\ell mn},
\end{eqnarray}
where $f_{\ell mn}$ are the  oscillation frequency, and $\tau_{\ell mn}$ are related to the damping time.  It is noted that this template based on the ringdown QNM frequencies  can really generate a complete ringdown GW waveform in GR for which the odd and even parity perturbations have the same spectrum. But in scalar-tensor theory  the isospectrality was found to be broken \cite{Li:2023ulk}, and other polarization modes, such as longitudinal and breathing modes, can  arise due to the scalar field \cite {Eardley:1973br}. Therefore, the template we utilize here is not realistic to generate a complete ringdown waveform in this specific Horndeski theory. Nevertheless, since the odd perturbation is decoupled from the scalar field, so here we assume the related ringdown GW can be treated separately and described by the aforementioned template for a preliminary study. Similar assumption was also taken in \cite{Lahoz:2023csk}. To begin with the error estimation under the detection of above waveform \eqref{eq:signal}, we need to define a useful inner product$(\cdot|\cdot)$, weighted by a noise spectral density $S_h(v)$ of a detector, between the signals $h_1(t)$ and $h_2(t)$:
\begin{eqnarray}\label{eq:signal inner product}
    (h_1|h_2)\equiv 2\int^{\infty}_{0} \frac{\tilde{h}^{*}_1\tilde{h}_2+ \tilde{h}^{*}_2\tilde{h}_1}{S_h(v)}dv,
\end{eqnarray}
with $\tilde{h}(v)$ defined by the Fourier transformation of the waveform $h(t)$ . Therefore, the signal-to-noise-ratio (SNR) can now be defined as the self-inner product of a single waveform
\begin{eqnarray}\label{eq:SNR}
    \rho^2=(h|h)= 4\int^{\infty}_{0} \frac{\tilde{h}^{*}(v)\tilde{h}(v)}{S_h(v)}dv.
\end{eqnarray}
Then, for a given parameter space $\{\theta_a\}$ associated with a template signal, one can estimate their errors with respect to measured signals by calculating the FIM 
\begin{eqnarray}\label{eq:fisher}
\Gamma_{ab}\equiv (\partial_a h| \partial_b h)\quad \text{with}\quad \partial_a\equiv\frac{\partial}{\partial\theta_{a}}.
\end{eqnarray}
Via inverting the FIM one obtains the covariance matrix and root-mean-square error for a single parameter
\begin{eqnarray}\label{eq:covariance}
    \Sigma_{ab}\equiv(\Gamma^{-1})_{ab}=\braket{\Delta\theta_a\Delta\theta_b},~~~~
    \sigma_{a}=\sqrt{\Sigma_{aa}}=\sqrt{\langle (\Delta\theta_a)^2 \rangle}.
\end{eqnarray}

For a preliminary work to perform parameter estimation, one usually impose some simplifying assumptions \cite{Berti:2005ys}:

(i). The wavelength of GWs is much larger than the arm length of interferometer, therefore, the spacial dependence of  the signals can be neglected and hence the product of the pattern and spin-weighted functions appear in the angle averages as $\braket{F^2_{+}}=\braket{F^2_{\times}}=1/5$, $\braket{F_{\times}F_{+}}=0$ and $\braket{|S_{\ell mn}|^2}=1/4\pi$.

(ii). Assume that the noise can be approximately seen as a constant over the bandwidth of the signals, so that we have $\sigma^2_{\text{error}} \sim S_h$ and $\rho^2_{\text{SNR}} \sim S^{-1}_h$. So a natural way to get rid of the dependence of this constant noise is to multiply the error by SNR, i.e. $\rho_{\text{SNR}} \,\sigma_{\text{error}}$, which shall then represent a generic error re-scaled by SNR.

(iii). To further simplify the expression for SNR and error, one can define 
\begin{eqnarray}
    A^{\times}_{\ell mn}&=&A^{+}_{\ell mn}N_{\times},\\
    \phi^{\times}_{\ell mn}&=&\phi^{+}_{\ell mn}+\phi^{0}_{\ell mn},
\end{eqnarray}
where the constant phase $\phi^{0}_{\ell mn}$ and the numerical factor $N_{\times}$ can be fixed because they are addressed to be less dependent on the error \cite{Berti:2005ys}.

We shall then concentrate our interest on the measurement error of the hair parameter $\tilde{Q}$ encoded within the QNMs frequencies $\omega_{QNM}$ such that we have $f_{\ell mn}(\tilde{Q})$ and $\mathcal{Q}_{\ell mn}(\tilde{Q})$. Since we are focusing on constraining a single parameter, so the Fisher matrix reduces to a ``Fisher scalar'' $\Gamma_{\tilde{Q}}=(\partial_{\tilde{Q}}h|\partial_{\tilde{Q}}h)$. Recalling the aforementioned assumptions,  the derivative on the waveform, $\partial_{\tilde{Q}}h$, is reflected in the derivatives on $\partial_{\tilde{Q}}f_{\ell mn}$ and $\partial_{\tilde{Q}}\mathcal{Q}_{\ell mn}$. Since our QNMs frequencies related to $\{f_{\ell mn},\mathcal{Q}_{\ell mn}\}$ are obtained discretely, we shall do these derivatives numerically in computing the Fisher scalar as well as the errors of the hair parameter. Consequently, for a fixed  hair parameter $\tilde{Q}_0$, the rescaled error $\sigma_{\tilde{Q_0}}$ defined in this way describes the deviation from  its  best estimated value by the detectors. We shall evaluate the rescaled error of the hair parameter by considering the signals composing of a single mode and two modes, respectively.

\subsection{Single-mode parameter estimation}
Since our black hole is static and spherically symmetric, the odd gravitational perturbation equation here  is independent of the azmuthal $m$ \cite{Ganguly:2017ort}, so we obviate  $m$ and consider the detecting signals are contributed only by a single mode with  
\begin{eqnarray}\label{eq:strain}
 h^{+}&=& h^{+}_{n\ell} = A^{+}e^{-\frac{\pi t f_{n\ell}}{\mathcal{Q}_{ n \ell}}}S_{n\ell}\;\cos(\phi^{+}+2 \pi t f_{n\ell}),\\
 h^{\times}&=& h^{\times}_{n\ell} =A^{+}N_{\times}e^{-\frac{\pi t f_{n\ell}}{\mathcal{Q}_{n\ell}}}S_{n\ell}\;\sin(\phi^{+}+\phi^{0}+2 \pi t f_{n\ell}).
\end{eqnarray}
Here we set $A^{+}=\frac{A^2}{1+N_{\times}^2}$, and $\{N_{\times}=1,\phi^{+}=\phi^{0}=0\}$ for simplicity, and apply the formulas (\ref{eq:signal inner product}--\ref{eq:covariance}) laid out above. Then the error (re-scaled by SNR) for a single-mode signal reads
\begin{eqnarray}\label{eq:error-single}
    \rho^2 \sigma_{\tilde{Q}}^2 \equiv \rho^2 \sigma^2_{n \ell}(\tilde{Q}) =\frac{2 f_{n\ell}^2 \mathcal{Q}_{n\ell}^2}{ f_{n\ell}'^2( \mathcal{Q}_{n\ell}^2+4 \mathcal{Q}_{n\ell}^4)-2 f_{n\ell} \mathcal{Q}_{n\ell} f_{n\ell}' \mathcal{Q}_{n\ell}'+f_{n\ell}^2 \mathcal{Q}_{n\ell}'^2},
\end{eqnarray}
 where the prime represents the derivative with respect to $\tilde{Q}$. 

\begin{figure*}[htbp]
\includegraphics[height=6cm]{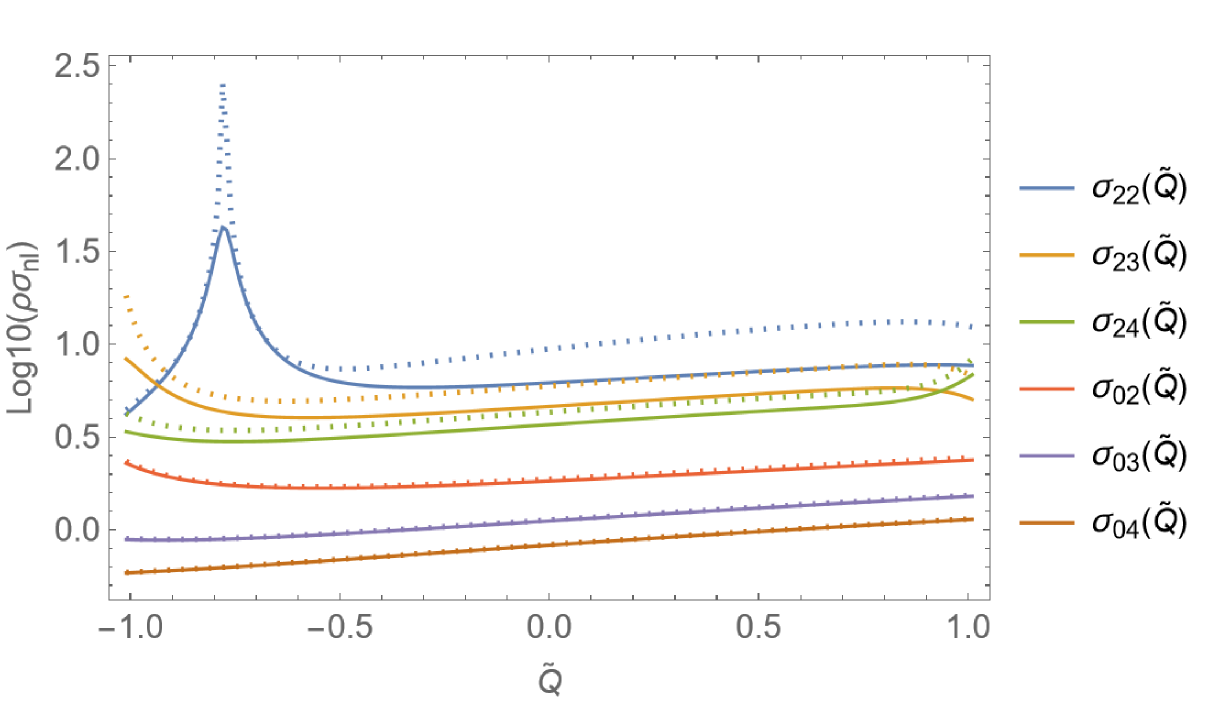}\hspace{0.1cm}
\caption{\label{error one mode} The SNR-scaled error for single-mode $h_{n\ell}$ measurements of hair parameter $\tilde{Q}$, with various modes. The dotted curves are the errors from the large-$\mathcal{Q}$ approximation.}
\end{figure*}

 As it was addressed in \cite{Vallisneri:2007ev} that $\Gamma^{-1}$ corresponds to the Cramer-Rao bound, which is a lower bound for the variance error of  any unbiased parameter. So in our case the error $\rho \sigma_{n \ell}(\tilde{Q})$ possesses the similar capability that describes a lower bound variance (re-scaled by SNR) on determining the source parameter $\tilde{Q}$. We compute the tendency of error as a function of hair parameter $\tilde{Q}$ for different single mode $h_{n \ell}$. As shown in FIG. \ref{error one mode}, each curve describes a scaled lowest variance, which would be the best extent where the detector can constrain the hair parameter in our setup. It can be clearly seen that the error bound of the hair parameter becomes deeper as $\ell$ increases but shallower as $n$ increases, while the effect from the hair parameter is relatively slight. In the figure, we also plot the errors for a large-$\mathcal{Q}$ approximation $\rho^2 \sigma^2_{n \ell} \simeq \frac{ f_{n\ell}^2 }{2 f_{n\ell}'^2 \mathcal{Q}_{n\ell}^2}$ which is directly derived from \eqref{eq:error-single}. A direct comparison reveals that this approximation works better for the error estimation of the fundamental mode.
 
\subsection{Two-modes parameter estimation}
The waveform mixed two different modes could be more significant in fitting astrophysical gravitational waves sources and testing the general no-hair theorem \cite{Berti:2005ys, Berti:2006hb, Barack:2003fp}, 
so in this part we would like to discuss how the constraint response in considering the mixed waveforms. To model such signal, we consider the superposition of two QNMs with different $\ell$ and $n$, which  can be written as
\begin{eqnarray}
h^{+}&=&h^{+}_{a}+h^{+}_{b}=A^{+}_{a} e^{-\frac{\pi t f_{n\ell}}{\mathcal{Q}_{n\ell}}}S_{n\ell} \;\cos(\phi^{+}_{a}+2\pi t f_{n\ell})+(\ell \longleftrightarrow\ell',n\longleftrightarrow n',a\longleftrightarrow b),\\
h^{\times}&=&h^{\times}_{a}+h^{\times}_{b}=A^{+}_{a}N_{\times} e^{-\frac{\pi t f_{n\ell}}{\mathcal{Q}_{n\ell}}}S_{n\ell} \;\sin(\phi^{+}_{a}+\phi^{0}_{a}+2\pi t f_{n\ell})+(\ell \longleftrightarrow\ell',n\longleftrightarrow n',a\longleftrightarrow b),
\end{eqnarray}
where for simplicity we have labeled the mode $(\ell,n)$ as ``$a$'' and $(\ell',n')$ as ``$b$''. In dealing with the product of the angular scalar between two different modes, we follow the assumption from \cite{Berti:2005ys} that two close enough  QNMs directly gives
\begin{eqnarray}\label{eq:two-modes product of angular scalar}
 \int S^{*}_{n' \ell'}\; S_{n\ell}\; d\Omega \simeq \delta_{\ell'\ell}.
\end{eqnarray}
By setting $\phi^{+}_{a}=-\phi^{0}_{a}=\phi^{\times}_{b}=\phi^{0}_{b}=\frac{1}{2\pi}$,  $A^{+}_{a}=A_{a}$,  $A^{+}_{b}=A_{b}$ and  $N_{\times}=1$, we will obtain the SNR and the Fisher scalar as
\begin{eqnarray}
\rho^2&=&\rho^2_{a}+\rho^2_{b}+\delta_{\ell'\ell}\;\rho^2_{c},\label{eq:mixed SNR}\\
    \Gamma_{\tilde{Q}}&=&\Gamma^{(a)}_{\tilde{Q}}+\Gamma^{(b)}_{\tilde{Q}}+ \delta_{\ell'\ell}\;\Gamma^{(c)}_{\tilde{Q}},\label{eq:mixed error}
\end{eqnarray}
with
\begin{eqnarray}
    \rho^2_{a}=\frac{A^2_{a}\mathcal{Q}_{n\ell}(1+\cos(\frac{1}{\pi})+8\mathcal{Q}^2_{n\ell})}{\pi S_{h}f_{n\ell}(1+4\mathcal{Q}^2_{n\ell})},\; \rho^2_{b}=\frac{A^2_{b}\mathcal{Q}_{n'\ell'}(2+\cos(\frac{1}{\pi})-\cos(\frac{1}{2\pi})+8\mathcal{Q}^2_{n'\ell'})}{\pi S_{h}f_{n'\ell'}(1+4\mathcal{Q}^2_{n'\ell'})}.
\end{eqnarray}
Here $\{\rho_{a}, \Gamma^{(a)}_{\tilde{Q}}\}$ and $\{\rho_{b}, \Gamma^{(b)}_{\tilde{Q}}\}$ denote the SNR and the Fisher scalar contributed by strains $h_{a}$ and $h_{b}$, respectively, while $\{\rho_c, \Gamma^{(c)}_{\tilde{Q}}\}$ represent the interacting term contributed by the product of two modes. We shall not present their expressions in details because of their complexity. 

\begin{figure*}[htbp]
\includegraphics[height=3.8cm]{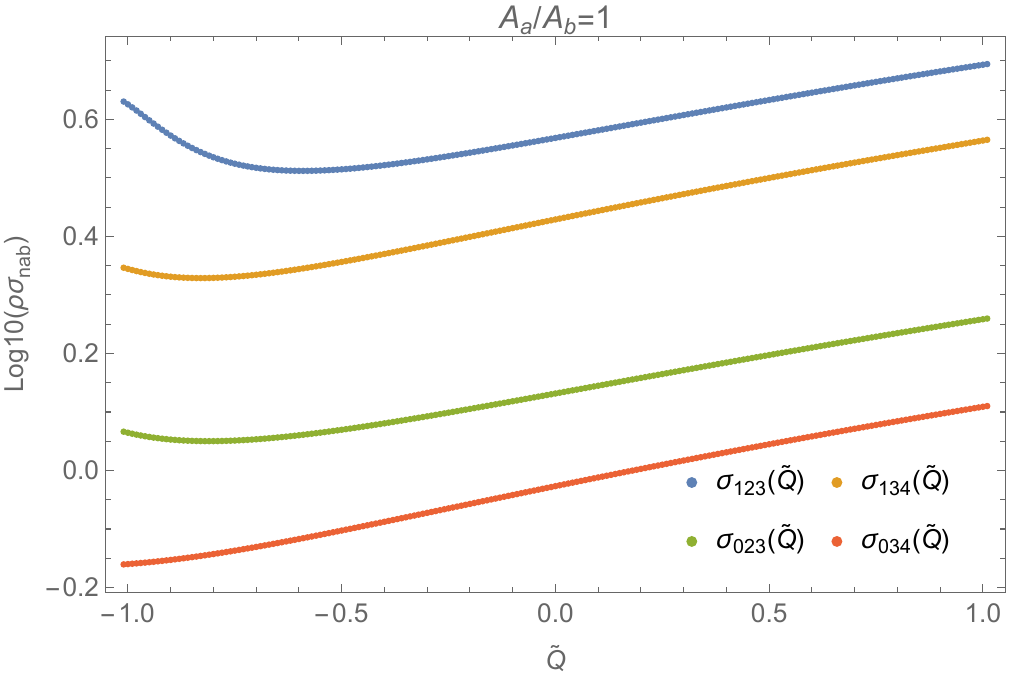}\hspace{0.1cm}
\includegraphics[height=3.8cm]{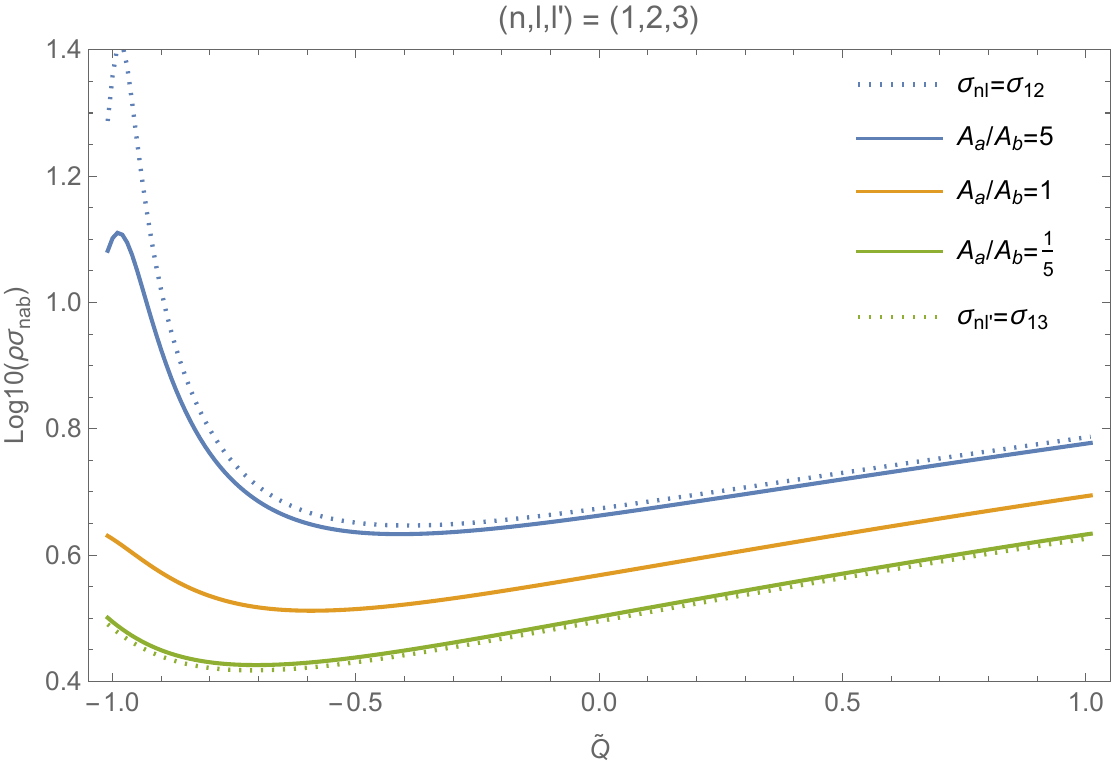}\hspace{0.1cm}
\includegraphics[height=3.8cm]{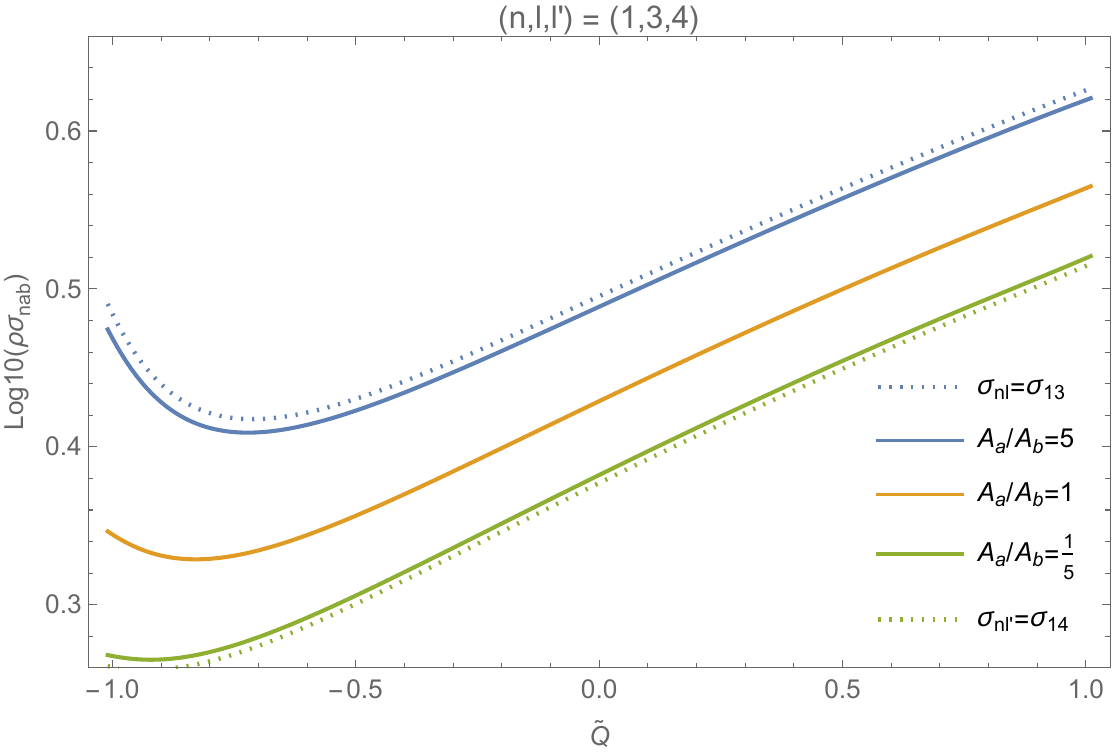}
\caption{\label{error two modes fix n} The SNR-scaled error for two-modes $\{h_a,h_b\}=\{h_{n\ell}, h_{n\ell'}\}$ measurements of hair parameter $\tilde{Q}$, the single modes $h_a$ and $h_b$ share a same overtone but different angular numbers. In the left plot, we have $(n,\ell,\ell')=(1,2,3)$ (blue), $(1,3,4)$ (orange), $(0,2,3)$ (green) and $(0,3,4)$ (red), respectively. The middle and right plots present the error for the mixed modes $(n,\ell,\ell')=(1,2,3)$ (middle) and $(1,3,4)$ (right),  but with different amplitude ratio $A_{a}/A_{b}$ between the two modes. The dotted curves denotes the single-mode errors.}
\end{figure*}

The appearance of a delta symbol in \eqref{eq:mixed SNR} and \eqref{eq:mixed error} makes the hair parameter estimation divide into two cases: 

(i) When $\ell'\neq \ell$ and $n'= n$, the cross term vanishes, so that the overall SNR and Fisher scalar become the sum in quadrature of two single modes and the scaled error are then expressed as
\begin{eqnarray}
    \rho^2\sigma^2_{\tilde{Q}}\equiv \rho^2\sigma^2_{nab}(\tilde{Q})=\frac{\rho^2_{a}+\rho^2_{b}}{\Gamma^{(a)}_{\tilde{Q}}+\Gamma^{(b)}_{\tilde{Q}}}.
\end{eqnarray}
Here $\sigma_{nab}$ denotes the variance of the mixed waveforms which have a same overtone but different angular numbers, for example, $\sigma_{023}$ means the mix of the waveforms $h_{02}$ and $h_{03}$.  In this case, similar to the one in the single mode estimation, the mixed mode with the same overtone but higher angular numbers exhibits lower error as shown in the left panel of FIG.\ref{error two modes fix n}. In order to discuss the subtle connections between the  single mode and mixed mode estimation, we plot the middle and right panels of FIG.\ref{error two modes fix n} from which we can read off the following features: (1) when $h_{b}=h_{n \ell'}$ dominates the mixed mode, i.e., $A_{a}/A_{b}<1$, the constraint becomes stronger  and quickly approaches the single mode $h_b$ itself (see the green curves). (2) when $h_{a}=h_{n \ell}$ dominates the mixed mode, i.e., $A_{a}/A_{b}>1$, the constraint becomes weaker. Therefore, our results show that the measurement  error from the mixed modes is rather sensitive to their relative amplitude.

\begin{figure*}[htbp]
\includegraphics[height=3.8cm]{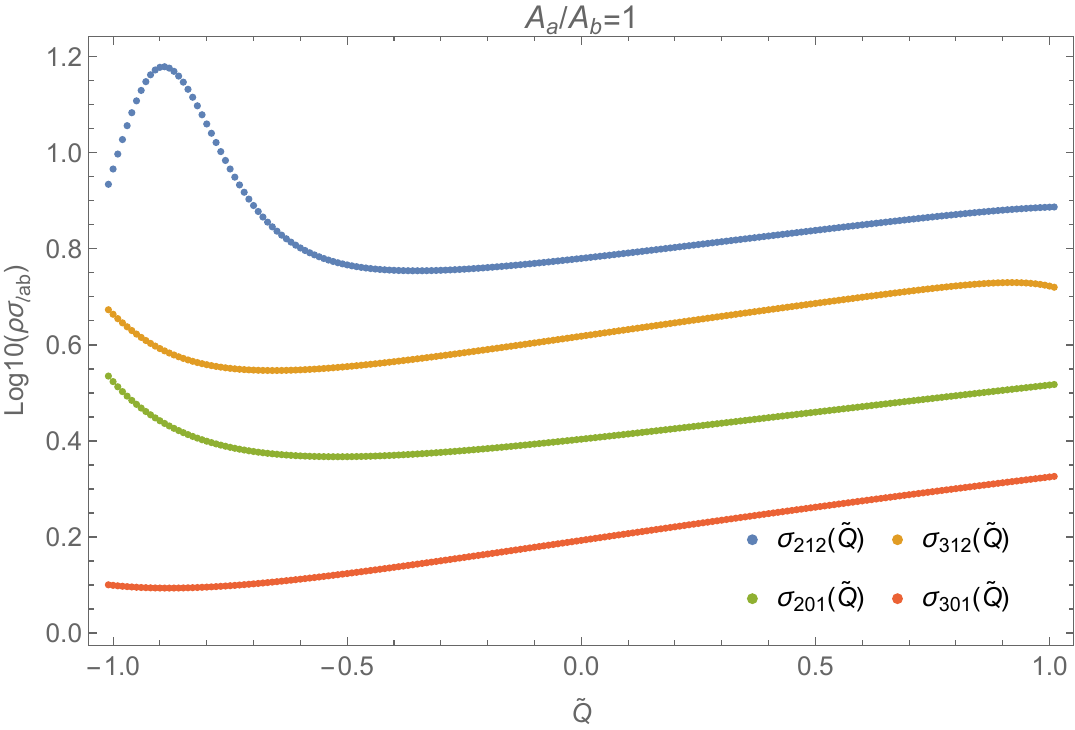}\hspace{0.1cm}
\includegraphics[height=3.8cm]{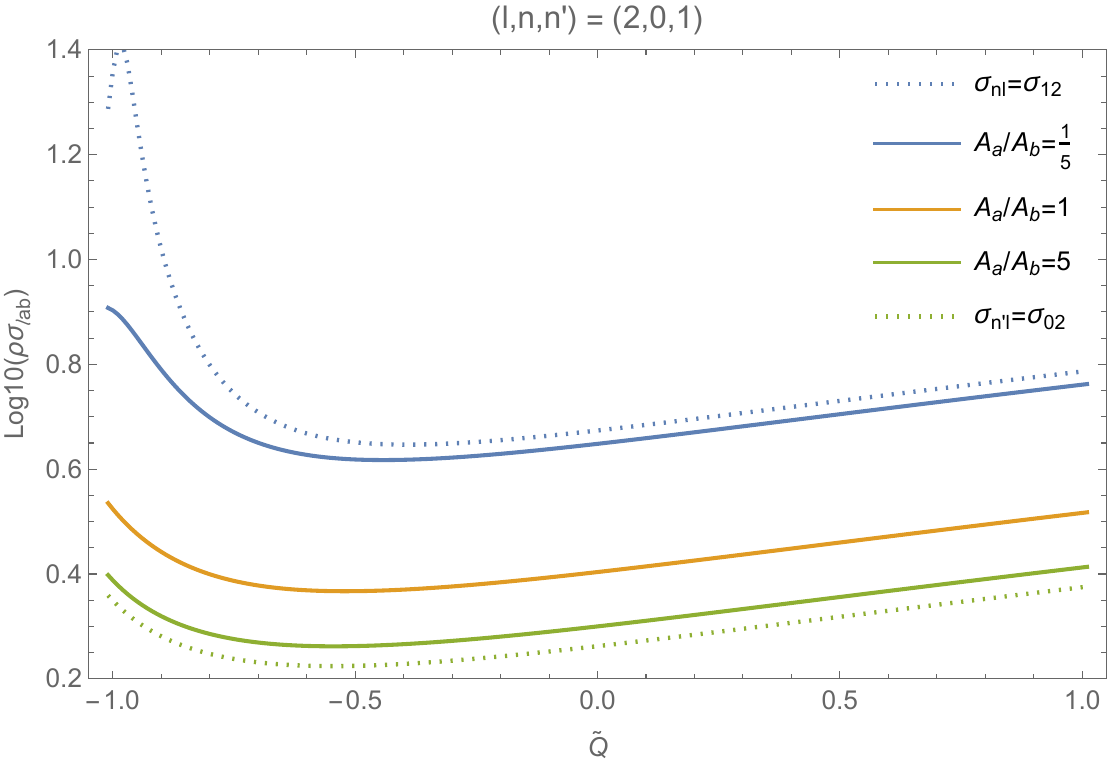}\hspace{0.1cm}
\includegraphics[height=3.8cm]{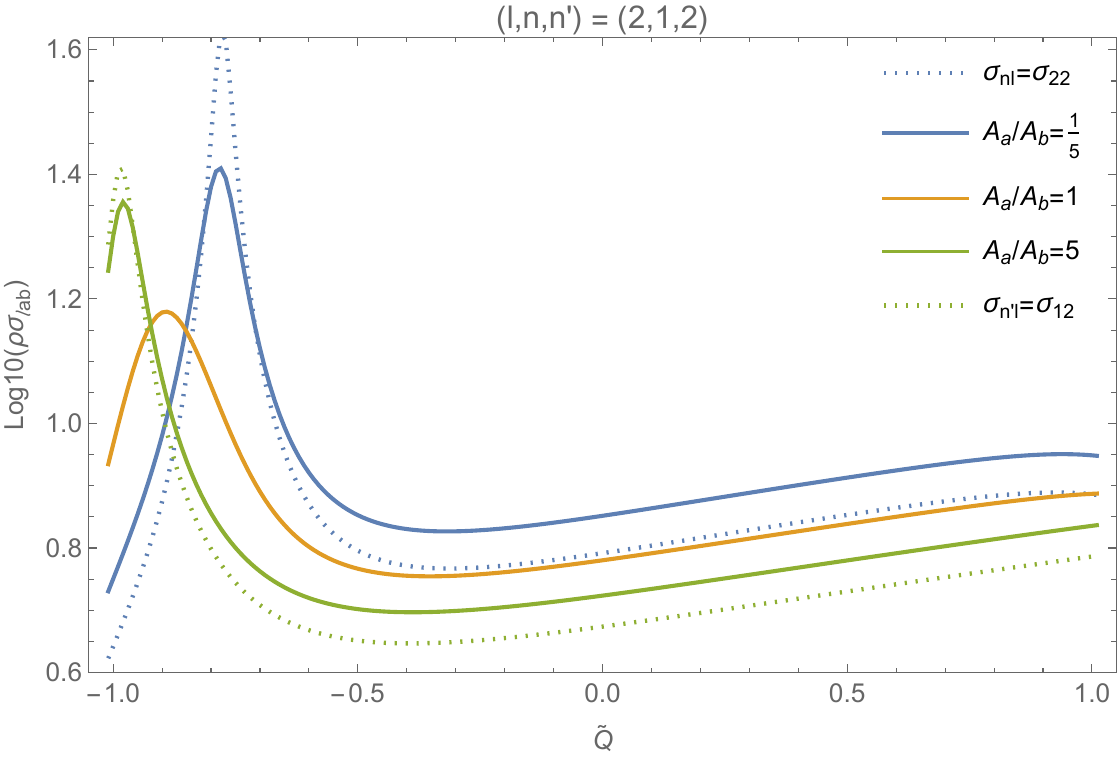}
\caption{\label{error two modes fix L}
The SNR-scaled error for two-modes $\{h_a,h_b\}=\{h_{n\ell}, h_{n'\ell}\}$ measurements of hair parameter $\tilde{Q}$, the single modes $h_a$ and $h_b$ share a same angular number but different overtones. In the left plot, we have $(\ell,n,n')=(2,1,2)$ (blue), $(3,1,2)$ (orange), $(2,0,1)$ (green), $(3,0,1)$ (red), respectively. In the middle and right plots, we select the error with $(\ell,n,n')=(2,0,1)$ (middle) and $(2,1,2)$ (right) and check the effect of the amplitude ratio $A_{a}/A_{b}=1/5$ of the two modes. The dotted curves are the single-mode errors.}
\end{figure*}

(ii) When $\ell'= \ell$ and $n'\neq n$, the product of the scalar spherical harmonics \eqref{eq:two-modes product of angular scalar} is approximated to unity, which introduce an additional term in overall SNR and Fisher scalar. Through straight algebraic calculations, we obtain  the scaled error in this case as
\begin{eqnarray}
    \rho^2\sigma^2_{\tilde{Q}}\equiv \rho^2\sigma^2_{\ell ab}(\tilde{Q})=\frac{\rho^2_{a}+\rho^2_{b}+\rho^2_{c}}{\Gamma^{(a)}_{\tilde{Q}}+\Gamma^{(b)}_{\tilde{Q}}+\Gamma^{(c)}_{\tilde{Q}}},
\end{eqnarray}
where $\sigma_{\ell ab}$ denotes the variance of the mixed mode with the same angular number but different overtones, for example, $\sigma_{201}$ mixes the waveforms $h_{02}$ and $h_{12}$. The Key results are depicted in the left panel of FIG.\ref{error two modes fix L},  showing that the mixed modes with same angular momentum but smaller overtones exhibit lower errors. To more explicitly see the mixing effects, we plot the middle and right panels. The middle and right panels indicate that when the fundamental  mode $h_a=h_{n\ell}$ dominates the mixing, {the overall error becomes smaller (see the green curves in middle and right panels of FIG.\ref{error two modes fix L} ), but if the next-to-leading overtone mode $h_b=h_{n'\ell}$ dominates the mixed modes, the error becomes larger. In addition, due to the existence of the interacting term $\Gamma^{(c)}_{\tilde{Q}}$, we also observe intriguing phenomena that the detection potential from the mixed modes is not always better than that from the single mode, and it is significantly affected by the amplitude ratio $A_{a}/A_{b}$ of the mixing mode as well as the hair parameter.

\section{Conclusions and discussions}\label{sec:conclustion}
In this paper we have investigated how a specific Horndeski hair differ the QNMs of the black hole from GR in the context of gravitational odd perturbation, and intuitively analyzed the varying of QNMs with the hair parameter by scrutinizing the evolution of the odd perturbation in the time domain. Moreover, as a preliminary attempt, we used the yielding QNM data to model ringdown waveforms of the hairy black hole and discussed the detectability of the hair parameter, which was estimated from the Fisher information matrix for a single parameter. Our key findings are summarized as follows:
\begin{itemize}
 \item Proper choices of auxiliary field and tortoise coordinate can refine the modified Regge-Wheeler equation into a standard one, such that the superluminal propagation problem in the odd perturbation  can be mathematically cured. Moreover, the refined equation has the same QNM spectrum as that for the original modified Regge-Wheeler equation. 

\item Comparing to the Shwarzschild black hole in GR, we found  that  the odd perturbation in the specific Horndeski gravity is significantly influenced by the Horndeski hair. We mainly analyzed the influence of the hair parameter on the QNMs frequencies. We found that in both frequency domain and time domain, the hairy black hole is dynamically stable under the odd perturbation. In particular, our results shew that when $\tilde{Q}$ becomes negative enough, the leading mode becomes  $\ell>2$ mode instead of $\ell=2$ mode, and also in this case the lifetime of the odd perturbation becomes the shortest one among all considered perturbations including massless scalar, Dirac, electromagnetic field perturbations. We found that these abnormal effects can be traced to the mismatch between the null geodesic analysis and the gravitational odd perturbation in the eikonal limit. A deeper physical explanation deserves further study.

\item In the context of parameter estimation by the ringdown waveform, we focused on the single parameter estimation of the Horndeski hair. Our results shew that the error bound of the mixed modes appeared in the average of the sum of corresponding single-mode errors. The ringdown observation would have a better constraint on the hair parameter when the waveforms carry the larger $\ell$ QNMs with smaller node $n$. The Horndeski black hole with negative $\tilde{Q}$ was found to have a better constraint comparing to the cases with positive hairy charge. 
Furthermore, FIG.\ref{error one mode}-FIG.\ref{error two modes fix L} indicate that the rescaled measurement error of the hair parameter is of order $\mathcal{O}(1-10)$. It means that the order of magnitude of the error on $\tilde{Q}$ for different GW detectors is around $\mathcal{O}(1-10)/\rho$ where $\rho$ is the Ringdown SNR for the  corresponding  detector. It is also worth noting that toward the end of this work, we became aware of a recent paper \cite{Sirera:2024ghv} in which the authors forecast how GW experiments could constrain the scalar parameter of hairy black hole using only the fundamental $\ell=2$ ringdown mode extracted by the WKB method in scalar-tensor theory.

Based on our present findings on the QNMs of gravitational odd perturbation and the error estimation, a series of further work can be proceeded to make the studies more physically realistic. Firstly, as an extension of the current work in the ringdown observation, a more comprehensive study on the parameter estimations, considering multi-parameters in the calculations of Fisher information matrix \cite{Berti:2005ys,Berti:2006hb,Berti:2009kk}, will be a natural next step. We believe the covariance of general parameters (for instance, spins, masses, redshift, etc.) will produce more information about the detectability of a hairy black hole. Secondly, it is interesting to fit the ringdown wave form more realistically by imposing the non-linear modification \cite{Qiu:2023lwo} on the overtone model or considering the non-linear modes \cite{Khera:2023oyf} in constructing the wave forms. Thirdly, other polarizations of gravitational waves modes may carry more interesting property on the observation of black hole spectroscopy (see for example \cite{Barbeau:2007qi,Rakhmanov:2004eh,Nishizawa:2009bf,Gong:2017bru}), and the effect of Horndeski hair on the GWs in EMRI system could be complementary to our current study. Those are worth to be studied for future work. The last but not the least, the extension of these studies to the gravitational even perturbation in the Horndeski gravity would  be another challenging direction, which is significantly required for a complete ringdown waveform construction. Though  the even gravitational perturbation always couples  with the scalar field in scalar tensor gravity, a method for the computation of QNMs frequencies in this coupled case was proposed in  \cite{Langlois:2021aji} and further applied in \cite{Roussille:2023sdr,Roussille:2024xwr}. The studies on all these topics could be significant for our understanding on the no-hair theorem and its testing via GW experiments. 
\end{itemize}

\begin{acknowledgments}
This work is partly supported by Natural Science Foundation of China under Grants No. 12375054 and the Postgraduate Research \& Practice Innovation Program of Jiangsu Province under Grant No. KYCX23\_3501.
\end{acknowledgments}

\bibliography{ref}
\bibliographystyle{utphys}

\end{document}